\documentclass[useAMS,usenatbib]{mn2e}
\usepackage{graphicx}
\def\aj{AJ}

\def\apj{ApJ}

\def\aap{A\&A}

\def\aaps{{A\&AS}}

\def\mnras{MNRAS}

\newcommand{\beq}{\begin{equation}}
\newcommand{\eeq}{\end{equation}}

\title[ ]{Multi-scale morphology of the galaxy distribution}
\author[ ]{Enn Saar$^{1}$, Vicent J. Mart\'{\i}nez$^{2}$,
Jean-Luc Starck$^{3}$ and David  L. Donoho$^{4}$\\
$^{1}$Tartu Observatoorium, T\~oravere, 61602, Estonia\\
$^{2}$ Observatori Astron\`omic, Universitat de Val\`encia, Apartat
de Correus 22085, E-46071 Val\`encia, Spain \\
$^{3}$ CEA-Saclay, DAPNIA/SEDI-SAP, Service d'Astrophysique, F-91191 Gif
sur Yvette, France \\
$^{4}$ Department of Statistics, Stanford University, Sequoia Hall,
Stanford, CA 94305, USA}
\begin{document}

\pagerange{\pageref{firstpage}--\pageref{lastpage}} \pubyear{2002}

\maketitle

\label{firstpage}

\begin{abstract}
Many statistical methods have been proposed in the last years for
analyzing the spatial distribution of galaxies. Very few of them,
however, can handle properly the border effects of complex
observational sample volumes. In this paper, we first show how to
calculate the Minkowski Functionals (MF) taking into account these
border effects. Then we present a multiscale extension of the MF
which gives us more information about how the galaxies are spatially
distributed. A range of examples using Gaussian random fields
illustrate the results. Finally we have applied the Multiscale
Minkowski Functionals (MMF) to the 2dF Galaxy Redshift Survey data.
The MMF clearly indicates an evolution of morphology with scale. We
also compare the 2dF real catalog with mock catalogs and found that
$\Lambda$CDM simulations roughly fit the data, except at the finest
scale.
\end{abstract}

\begin{keywords}
methods: data analysis -- methods: statistical -- large-scale structure of universe
\end{keywords}
\section{Introduction\label{sec:intro}}

One of the main tenets of the present inflationary paradigm is the
assumption of Gaussianity for the primordial density perturbations.
This postulate forms the basis of present theories of formation and
evolution of large-scale structure in the universe, and of its
subsequent analysis. But it remains a hypothesis that needs to be
checked.

The most straightforward way to do that would be to follow the
definition of Gaussian random fields (see, e.g, \citet{adler}) --
their one-point probability distribution and all many-point joint
probability distributions of field amplitudes have to be Gaussian.
This is clearly a too formidable task. Another way is to check the
relationships between the correlation functions and power spectra of
different orders, which are well-defined for Gaussian random fields.
This approach is frequently used (see, e.g., a review in
\citet{martinezsaar}). A third method is to study the morphology of
the cosmological (density) fields. One approach to the morphological
description relies on the so-called Minkowski functionals and is
complementary to the moment-based methods because these functionals depend
on moments of all orders. This procedure has been usually referred to
as topological analysis. It has a quite long history
already, starting with the seminal paper by \citet{gott86}, that
deals with the genus, a quantity closely related to one of the four
Minkowski functionals. The approach in the present paper lies within
this latter framework; we describe it in detail in
Sec.~\ref{sec:morph}.

There are two different possibilities to develop a morphological
analysis of galaxy catalogues based on Minkowski functionals. First,
we can dress all points (galaxies) with spheres of a given radius,
and study the morphology of the surface that is generated by the
convex union of these spheres, as a function of the radius, which
acts here as the diagnostic parameter. An appropriate theoretical
model to compare with in this case is a Poisson point process. On
the other hand, if we wish to study the morphology of the underlying
realization of a random field, we have to restore the (density)
field first, to choose an isodensity surface corresponding to a
given density threshold and to calculate its morphological
descriptors. In this approach the density threshold (or a related
quantity) acts as the diagnostic parameter. The theoretical
reference model is that of a Gaussian random field, and the crucial
point here is to properly choose a restoration method that
provides a smoothed underlying density field that should be fairly
sampled by the observed discrete point distribution.

Starting from the original paper on topology by \citet{gott86}, this
task has been typically done by smoothing the point distribution with a
Gaussian kernel. The choice of the optimal width of this kernel has
been widely discussed; it is usually taken close to the correlation
length of the point distribution. However, Gaussian smoothing is not
the best choice for morphological studies. As we have shown recently
\citep{mart05}, it tends to introduce additional Gaussian features
even for manifestly non-Gaussian density distributions. Minkowski
functionals are very sensitive to small density variations, and the
wings of Gaussian kernels could be wide enough to generate a
small-amplitude Gaussian ripple that is added to the true density
distribution. Such effect could be alleviated by using compact
adaptive smoothing kernels, as we show below.

\begin{figure}
\centering
\resizebox{.48\textwidth}{!}{\includegraphics*{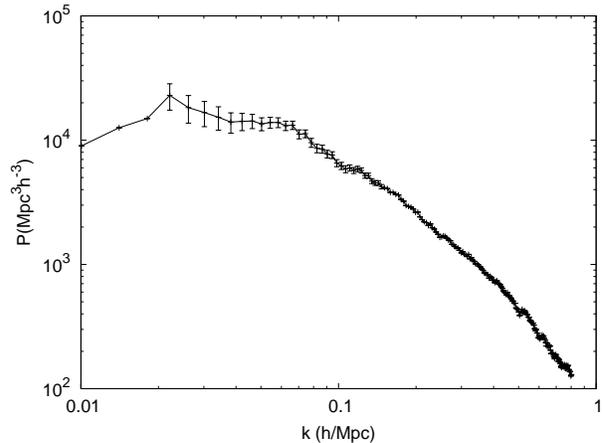}}
\caption{The matter density power spectrum for the 2dF
GRS. Data courtesy of W.~Percival and the 2dF GRS team.
\label{fig:2dfpk}}
\end{figure}

It is well known that large-scale cosmological fields have a
multi-scale structure. A good example is the density field; it
includes components that vary on widely different sales. The
amplitudes of these components can be characterized by the power
spectrum; the present determinations encompass the frequency
interval 0.01--0.8 $h$/Mpc,\footnote{As usual, $h$ is the present
Hubble parameter, measured in units of 100 km/sec/Mpc.} which
corresponds to the scale range from 8 to over 600 Mpc/$h$.
Fig.~\ref{fig:2dfpk} shows the power spectrum for the  2dF galaxy
redshift survey (2dFGRS).

As cosmological densities have many scales and widely varying
amplitudes, density restoration should be adaptive. Different
methods exist to adaptively smooth point distributions to estimate
from them the underlying density field. \citet{schaap00} have
introduced the Delaunay Tessellation Field Estimator (DTFS) which
adapts itself to the point configuration even when anisotropies are
present. The method starts by considering the Delaunay tessellation
of the point process, then we can estimate the density at those 
points using the contiguous Voronoi cells, and finally, we should
interpolate to obtain the density in the whole volume. Intricate 
point patterns have been successfully smoothed using this method 
and applications to particle
hydrodynamics provide good performance \citep{pelupessy03}.
\citet{ascasibar05} have recently introduced a novel technique based
on a different partition of the embedding space. These authors use
multidimensional binary trees to make the partition and latter 
apply adaptive kernels within the resulting cells. 
Finally, it is well known that wavelets provide a
localized (compact-kernel) adaptive restoration method
\citep{starck:book02}. We have applied, in a previous paper
\citep{mart05}, a wavelet based denoising technique to the 2dFGRS. 
As a result we found that the morphology
of the galaxy density distribution in the survey volume does not
follow a Gaussian pattern, in contrast to the usual results in which
deviations of Gaussianity are not clearly detected (see,
e.g., \citet{hoyle02} for the 2dFGRS and \citet{park05} for the
Sloan Digital Sky Survey (SDSS)).

By the way, adaptive density restoration methods are probably the
best for calculating partial Minkowski functionals, to describe the
morphology of single large-scale density enhancements (superclusters;
see, e.g., \cite{shandarin04}). Partial functionals can be used to characterize
the inner structure (clumpiness) and shapes (via shapefinders
\citep{sahni98}) of superclusters. As Minkowski functionals are
additive, partial functionals can be, in principle, combined to
obtain global Miknkowski functionals for the whole catalogue volume.
But if we want to check for non-Gaussianity, direct calculation of
global Minkowski functionals is more simple and straightforward.
When combining partial functionals, estimating the mean densities
and volume distributions for the full sample is a  difficult
problem.

Now, although a single adaptively found density distribution
represents the cosmological density field better, it could not be
the best tool for comparing theories with observations. Theories of
evolution of structure predict that Gaussianity of the original
density distribution is distorted during evolution, and this
distortion is scale-dependent. According to the present paradigm,
evolution of structure should proceed with different pace at
different scales. At smaller scales signatures of gravitational
dynamics should be seen, and traces of initial conditions could be
discovered at larger scales. In a single density field, containing
contributions from all scales, these effects are mixed. Thus, a
natural way to study cosmological density fields is the multi-scale
approach, scale by scale. This has been done in the past by using a
series of kernels of different width \citep{park05}, but this method
retains considerable low-frequency overlap. A better way is to
decompose the density field into different frequency (scale)
subbands, and to study each subband separately.

The simplest idea of separation of scales by using different Fourier
modes does not work well, at least for morphological studies
(studies of shapes and texture). Describing texture requires
knowledge of positions, but Fourier modes do not have positions,
their position is the whole sample space. A similar weakness, to a
smaller extent, is shared by discrete orthogonal wavelet expansions
-- their localisation properties are better, but vary with scale,
and large-scale modes remain badly localised.

This leads to the conclusion that the natural candidates for scale
separation are shift-invariant wavelet systems, where wavelet amplitudes
of all scales are calculated for each point of the coordinate grid. These wavelet
decompositions are redundant -- each subband has the same data
volume as the original data. 

For such a scheme, direct calculation of
low-frequency subbands would require convolution with wide wavelet
profiles, that could be numerically expensive. A way out is the
\emph{\`a trous} (with holes) trick, where convolution kernels for
the next dyadic scale are obtained by inserting zeros between the
elements of the original kernel. In this way, the total number of
non-zero elements is the same for all kernels, and all wavelet
transforms are equally fast.

Now, as usual with wavelets, there is considerable freedom in
choosing the wavelet kernel. A particularly useful choise is the
wavelet based on a $B_3$ spline scaling function (see Appendix~A).
In this case, the original data can be reconstructed as a simnple
sum of subbands, without extra weights, so the subband decomposition
is the most natural.

In the present paper we combine the \`a trous representation of
density fields with a grid-based algorithm to calculate the
Minkowski functionals (MF for short), and apply it to the
\mbox{2dFGRS} data.

Section~\ref{sec:morph} describes how to compute the MF for complex
data volumes and how to extend such an approach in a multiscale
framework using the wavelet transform. Section~\ref{sec:data}
describes our observational data while section~\ref{sec:gauss}
evaluates the multiscale MF on Gaussian random field realizations
for which the analytical results are known.
Section~\ref{sec:datamorph}
presents our results for the 2dFGRS data. These results are compared
with the multiscale MF calculated for 22 mock surveys and about 100
Monte-Carlo simulations of Gaussian random fields. We list the
conclusions in section~\ref{sec:concl}.

\section{Morphological analysis}
\label{sec:morph}

\subsection{Definition}
An elegant description of morphological characteristics of density
fields is given by Minkowski functionals \citep{cf:mecke94}. These
functionals provide a complete family of morphological measures. In
fact all additive, motion invariant and conditionally continuous
\footnote{The functionals are required to be continuous only for
compact convex sets; we can always represent any hypersurface as
unions of such sets.} functionals defined for any hypersurface are
linear combinations of its Minkowski functionals.

The Minkowski functionals describe the morphology of iso-density
surfaces \citep{minkowski,tomita}, and depend thus on the specific
density level (see \citet{sheth05} for a recent review). Of course,
when the original data are galaxy positions, the procedure chosen to
calculate densities (smoothing) will also determine the result
\citep{mart05}. Generally, convolution of the data with a Gaussian
kernel is applied to obtain a continuous density field from the
point distribution. An alternative approach starts from the point
field, decorating the points with spheres of the same radius, and
studying the morphology of the resulting surface
\citep{ker96var,ker2}. This approach does not refer to a density; we
cannot use that for the present study.

The Minkowski functionals are defined as follows. Consider an
excursion set $F_{\phi_0}$ of a field $\phi(\mathbf{x})$ in 3-D (the
set of all points where $\phi(\mathbf{x}\ge\phi_0$). Then, the first
Minkowski functional (the volume functional) is the volume of the
excursion set:
\[
V_0(\phi_0)=\int_{F_{\phi_0}}d^3x.
\]
The second MF is proportional to the surface area
of the boundary $\delta F_\phi$ of the excursion set:
\[
V_1(\phi_0)=\frac16\int_{\delta F_{\phi_0}}dS(\mathbf{x}).
\]
The third MF is proportional to the integrated mean curvature
of the boundary:
\[
V_2(\phi_0)=\frac1{6\pi}\int_{\delta F_{\phi_0}}
    \left(\frac1{R_1(\mathbf{x})}+\frac1{R_2(\mathbf{x})}\right)dS(\mathbf{x}),
\]
where $R_1$ and $R_2$ are the principal curvatures of the boundary.
The fourth Minkowski functional is proportional to the integrated
Gaussian curvature (the Euler characteristic) of the boundary:
\[
V_3(\phi_0)=\frac1{4\pi}\int_{\delta F_{\phi_0}}
    \frac1{R_1(\mathbf{x})R_2(\mathbf{x})}dS(\mathbf{x}).
\]
The last MF is simply related to other known morphological
quantities
\[
V_3=\chi=\frac12(1-G),
\]
where $\chi$ is the Euler characteristic and $G$ is the
topological genus, widely used in the past study of cosmological
density distributions. The
functional $V_3$ is a bit more comfortable to use -- it is additive,
while $G$ is not, and it gives just twice the number of isolated
balls (or holes).
Instead of the functionals, their spatial densities $V_i$ are
frequently used:
\[
v_i(f)=V_i(f)/V, \quad i=0,\dots,3,
\]
where $V$ is the total sample volume. The densities allow us to
compare the morphology of different data samples.

The original argument of the functionals, the density level
$\rho_0$, can have different amplitudes for different fields, and
the functionals are difficult to compare. Because of that,
normalised arguments are usually used; the simplest one is the
volume fraction $f_v$, the ratio of the volume of the excursion set
to the total volume of the region where the density is defined.
Another, similar argument is the mass ratio $f_m$, which is very
useful for real, positive density fields, but is cumbersome to apply
for realizations of Gaussian fields, where the density may be
negative. The most widely used argument is the Gaussianized volume
fraction $\nu$, defined as \beq \label{nu}
f_v=\frac1{\sqrt{2\pi}}\int_\nu^\infty\exp(-t^2/2)\,dt. \eeq For a
Gaussian random field, $\nu$ is the density deviation from the mean,
divided by the standard deviation. This argument was introduced
already by \citet{gott86}), in order to eliminate the first trivial
effect of gravitational clustering, the deviation of the 1-point pdf
from the (supposedly) Gaussian initial pdf. Notice that using this
argument, the first Minkowski functional is trivially Gaussian by
definition.

All the Minkowski functionals have analytic expressions for
iso-density slices of realizations of Gaussian random fields.
For three-dimensional space they are \citep{tomita}:
\begin{eqnarray}
\label{gaussv}
v_0&=&\frac12-\frac12\Phi\left(\frac{\nu}{\sqrt2}\right),\\
v_1&=&\frac23\frac{\lambda}{\sqrt{2\pi}}\exp\left(-\frac{\nu^2}2\right),\\
v_2&=&\frac23\frac{\lambda^2}{\sqrt{2\pi}}\nu\exp\left(-\frac{\nu^2}2\right),\\
v_3&=&\frac{\lambda^3}{\sqrt{2\pi}}(\nu^2-1)\exp\left(-\frac{\nu^2}2\right),
\end{eqnarray}
where $\Phi(\cdot)$ is the Gaussian error integral, and $\lambda$
is determined by the correlation function $\xi(r)$ of the field:
\beq
\label{lambda}
\lambda^2=\frac1{2\pi}\frac{\displaystyle\xi''(0)}{\displaystyle\xi(0)}.
\eeq

\subsection{Numerical algorithms}

Several algorithms are used to calculate the Minkowski functionals
for a given density field and a given density threshold.  We can
either try to follow exactly the geometry of the iso-density
surface, e.g., using triangulation \citep{surfgen}, or to
approximate the excursion set on a simple cubic lattice. The
algorithm that was proposed first by \citet{gott86}, uses a
decomposition of the field into filled and empty cells, and another
popular algorithm \citep{coles96} uses a grid-valued density
distribution. The lattice-based algorithms are simpler and faster,
but not as accurate as the triangulation codes.

We use a simple grid-based algorithm, that makes use of integral
geometry (Crofton's intersection formula, see \citet{jens97}). We
find the density thresholds for given filling fractions by sorting
the grid densities, first. Vertices with higher densities than the
threshold form the excursion set. This set is characterised by its
basic sets of different dimensions -- points (vertices), edges
formed by two neighbouring points, squares (faces) formed by four
edges, and cubes formed by six faces. The algorithm counts the
numbers of elements of all basic sets, and finds the values of the
Minkowski functionals as
\begin{eqnarray}
\label{crofton}
V_0(f)&=&a^3N_3,\nonumber\\
V_1(f)&=&a^2\left(\frac29N_2(f)-\frac23N_3(f)\right),\nonumber\\
V_2(f)&=&a\left(\frac29N_1(f)-\frac49N_2(f)+\frac23N_3(f)\right),\nonumber\\
V_3(f)&=&N_0(f)-N_1(f)+N_2(f)-N_3(f),
\end{eqnarray}
where $a$ is the grid step, $f$ is the filling factor, $N_0$ is the
number of vertices, $N_1$ is the number of edges, $N_2$ is the
number of squares (faces), and $N_3$ is the number of basic cubes in
the excursion set for a given filling factor (density threshold).
The formula (\ref{crofton}) was first used in
cosmological studies by \citet{coles96}.

\subsection{Biases}
The algorithm described above is simple to program, and is very fast,
allowing the use of Monte-Carlo simulations for error estimation.

\begin{figure}
\centering
\resizebox{.48\textwidth}{!}{\includegraphics*{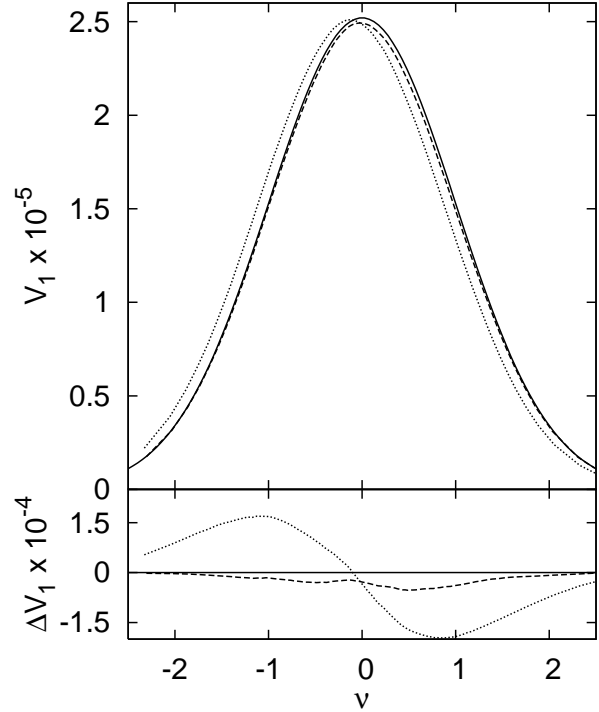}}
\caption{The $V_1$ functional for a realization of a Gaussian
random density field in a periodic $256^3$ cube (upper panel).
Full line shows the theoretical prediction for this realization,
dotted line -- the standard one-excursion-set estimate,
and dashed line -- the average of the functional over two excursion sets,
The lower panel shows the difference between the estimates and
the theoretical prediction; the lines encode the estimates as above.
\label{fig:v1shift}}
\end{figure}

However, it suffers from discreteness errors, which are not large,
but annoying, nevertheless. An example of that is given in
Fig.~\ref{fig:v1shift}, where we show the $V_1$ functional,
calculated by the above recipes for a periodic realization of a
Gaussian field (the dashed line).  As we see, it has a constant
shift in $\nu$ over the whole range. This shift is due to the fact
that when we approximate iso-density surfaces by a discrete grid,
the vertices that compose the surface lie in a range of densities
starting from the nominal one. This effect can easily be calculated
because this bias will show up as a constant shift in $\nu$ for a
Gaussian density field, as observed. Other functionals ($V_2$ and
$V_3$) suffer similar shifts, with smaller amplitudes, and these are
not easy to explain.

There is, fortunately, another and simple possibility to fight these
errors. The standard way is to approximate an iso-density surface by
the collection of vertices that have densities $\rho\ge\rho_l$,
where $\rho_l$ is the threshold density. But another surface, formed
by the vertices with $\rho<\rho_l$, is as good an approximation to
the iso-density surface as the first one. Thus, the natural way to
calculate the Minkowski functionals is to run the algorithm twice,
swapping the marks for the excursion set, and averaging the values
of the functionals obtained. The last step is justified, as the
Minkowski functionals are additive. The averaging rules are:
\begin{eqnarray*}
V_0&=&\left(V_0^{(1)}+V_{\mbox{\scriptsize tot}}-V_0^{(2)}\right)/2,\\
V_1&=&\left(V_1^{(1)}+V_1^{(2)}\right)/2,\\
V_2&=&\left(V_2^{(1)}-V_2^{(2)}\right)/2,\\
V_3&=&\left(V_3^{(1)}+V_3^{(2)}\right)/2,\\
\end{eqnarray*}
Here the upper indices $(1)$ and $(2)$ denote the original and
complementary excursion sets, respectively, and
$V_{\mbox{\scriptsize tot}}$ is the total number of grid cubes in
the data brick. The minus sign in the formula for the third
functional ($V_2$) accounts for the fact that the curvature of the
second surface is opposite to that of the first one.

\begin{figure}
\centering
\resizebox{.48\textwidth}{!}{\includegraphics*{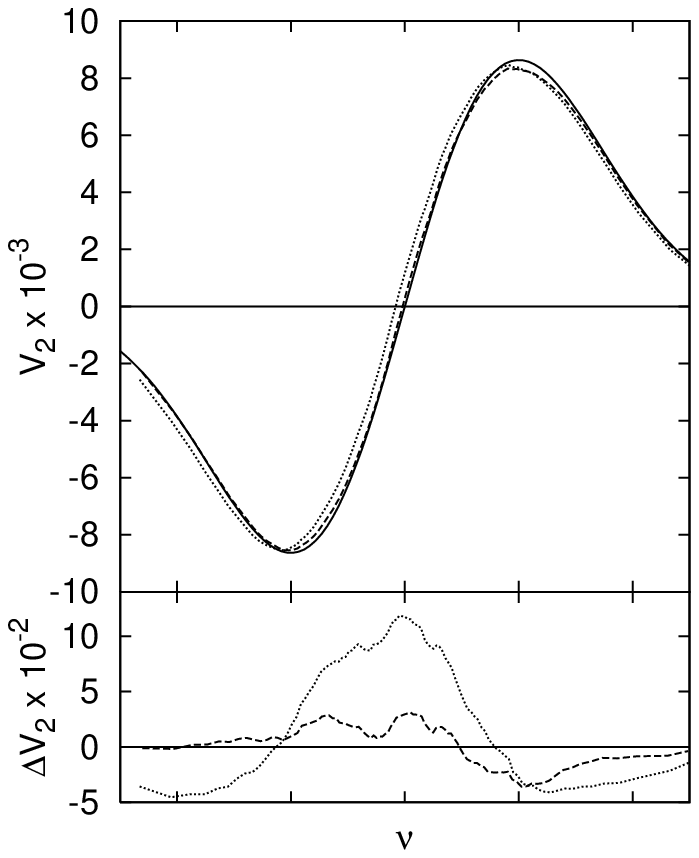}}\\
\resizebox{.48\textwidth}{!}{\includegraphics*{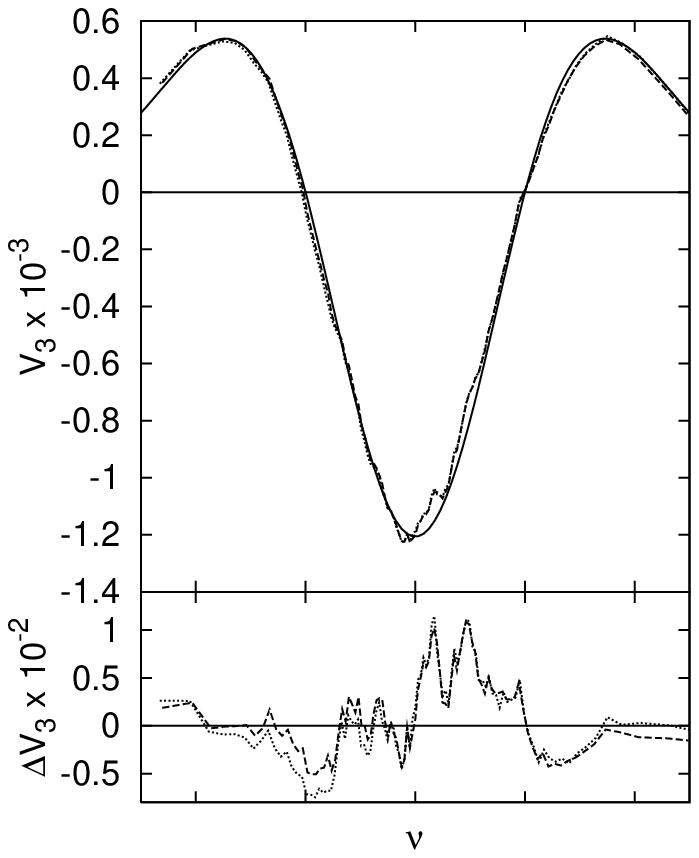}}\\
\caption{The $V_2$ (upper panels) and $V_3$ (lower panels)
functionals for a realization of a Gaussian
random density field in a periodic $256^3$ cube.
In the larger panels
full lines show the theoretical predictions for this realization,
dotted lines -- the standard one-excursion-set estimates,
and dashed lines -- the averages of the functionals over two excursion sets,
The smaller panels show the differences between the estimates and
the theoretical predictions; the lines encode the estimates as above.
\label{fig:v23shift}}
\end{figure}

The $V_1$ functional calculated this way is shown in
Fig.~\ref{fig:v1shift} by the full line; we can compare it with the
theoretical prediction for Gaussian fields ((\ref{gaussv}), the
dotted line). The coincidence of the two curves is very good; the
only slight deviation is at $\nu\approx0$, where the Gaussian
surface is more complex. We have to stress that the Gaussian curve
is not a fit; the parameter $\lambda$ that determines the amplitude
of the curve was found directly from the data, using the relations
$\xi(0)=\langle\rho^2\rangle$ and
$\xi''(0)=\langle\rho_{,i}^2\rangle$, where $\xi(r)$ is the
correlation function and $\rho_{,i}$ is the derivative of density at
a grid vertex in one of the coordinate directions. The good match of
these curves shows also that the Gaussian realization is good, which
is not simple to model. The averaging works as well for the two
other functionals; this is shown in Fig.~\ref{fig:v23shift}. There
are slight deviations from the theoretical curve for $V_2$ around
$\nu\approx1$ and for $V_3$ at $\nu\approx0$; these may be intrinsic
to the particular realization, as the number of 'resolution details'
diminishes when the order of the functional increases.

We also see that the higher the order, the closer are the
one- and two- excursion set estimates. So, even if we are interested
only in the topology of the density iso-surfaces, we should
correct for the border effects, all Minkowski functionals are used,
and it is important that they were unbiased.

There is a natural restriction on the grid steps  -- the grid has to
be fine enough to resolve the details of the density field. The
previous figures (Figs.~\ref{fig:v1shift}, \ref{fig:v23shift}) show
the Minkowski functionals obtained for the case of the Gaussian
field smoothed by a Gaussian filter with $\sigma=3$ grid steps,
being the smoothing radius $R\approx2\sigma=6$.

\subsection{Border corrections\label{subsec:bound}}
As we have seen above, we can obtain good estimates of the Minkowski
functionals for periodic fields. The real data, however, is always
spatially limited, and the limiting surfaces cut the iso-density
surface. An extremely valuable property of Minkowski functionals is
that such cuts can be corrected for. Let us assume that the data
region (window or mask) is big enough relative to the typical size
of details, so that one can consider the field inside the mask
homogeneous and isotropic. For this case, \citet{ker96var} show that
the observed Minkowski functionals for the masked iso-surface
$M_i(D\cap W)$ can be expressed as a combination of the true
functionals $M_i(D)$ and those of the mask $M_i(W)$: \beq
\label{kinform} M_i(D\cap W)=\frac1{\cal V}\sum_{j=0}^i
     \left(\begin{array}{c}i\\j\end{array}\right)
     M_j(D)M_{i-j}(W),
\eeq where $\cal V$ is the total volume inside the mask. Note that
the functionals $M_i$ differ from the usual $V_i$ by normalisation.
\citet{ker96var} derive the relation (\ref{kinform}) for a
collection of balls. Here we have applied it to iso-density surfaces
and for the true values of the functionals, we get \beq
\label{mcorr} \frac{M_i(D)}{\cal V}=\frac{M_i(D\cap W)}{M_0(W)}-
    \sum_{j=0}^{i-1}\left(\begin{array}{c}i\\j\end{array}\right)
    \frac{M_j(D)}{\cal V}\frac{M_{i-j}(W)}{M_0(W)}.
\eeq

The relation between $M_i$-s and the usual $V_i$-s is
\[
M_i=\frac{\omega_{d-i}}{\omega_d}V_i,
\]
where $\omega_j$ is the volume of a $j$-dimensional unit ball,
and $d$ is the dimension of the space.
For Minkowski functionals in three-dimensional space,
the explicit  relations are:
\[
M_0=V_0;\quad M_1=\frac34 V_1;\quad M_2=\frac3{2\pi}V_2;\quad M_3=\frac3{4\pi}V_3.
\]
Using (\ref{mcorr}) and replacing $M_i$-s by $V_i$, we arrive
at the following correction chain:
\begin{eqnarray}
\label{corrchain}
v_0(\nu)&=&\frac{V_0(\nu)}{V_0(W)}, \nonumber \\
v_1(\nu)&=&\frac{V_1(\nu)}{V_0(W)}-v_0(\nu)\frac{V_1(W)}{V_0(W)},\nonumber \\
v_2(\nu)&=&\frac{V_2(\nu)}{V_0(W)}-v_0(\nu)\frac{V_2(W)}{V_0(W)}
            -\frac{3\pi}{4}v_1(\nu)\frac{V_1(W)}{V_0(W)},\nonumber \\
v_3(\nu)&=&\frac{V_3(\nu)}{V_0(W)}-v_0(\nu)\frac{V_3(W)}{V_0(W)}
            -\frac92 v_1(\nu)\frac{V_2(W)}{V_0(W)} \nonumber \\
 & &
            -\frac92 v_2(\nu)\frac{V_1(W)}{V_0(W)}.\nonumber\\
\end{eqnarray}
Here $V_i(\nu)$ denote the observed (raw) values of Minkowski
functionals, and $v_i(\nu)$ denote the corrected densities.
\begin{figure*}
\centering
\resizebox{.32\textwidth}{!}{\includegraphics*{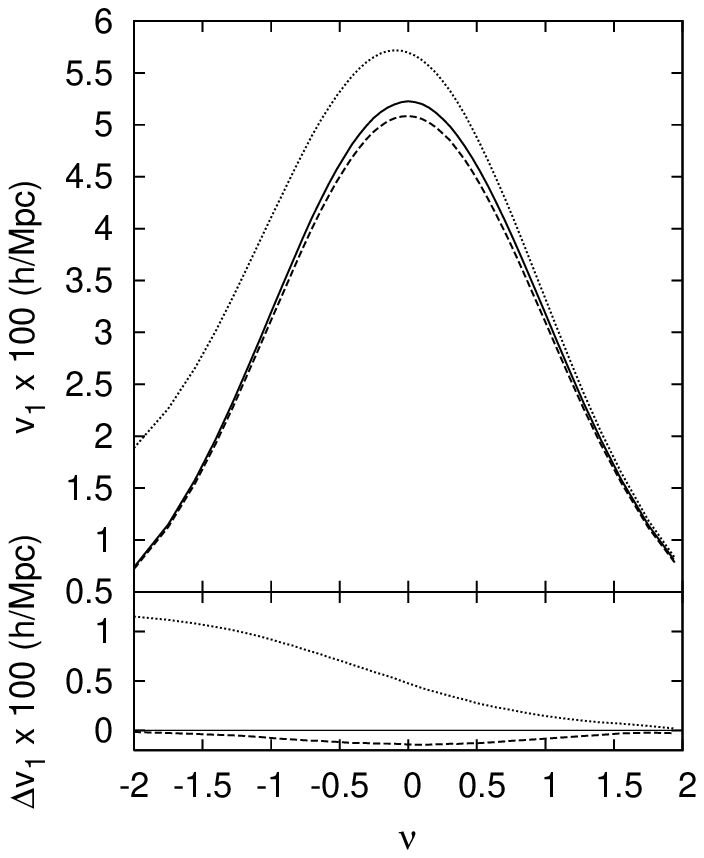}}
\resizebox{.32\textwidth}{!}{\includegraphics*{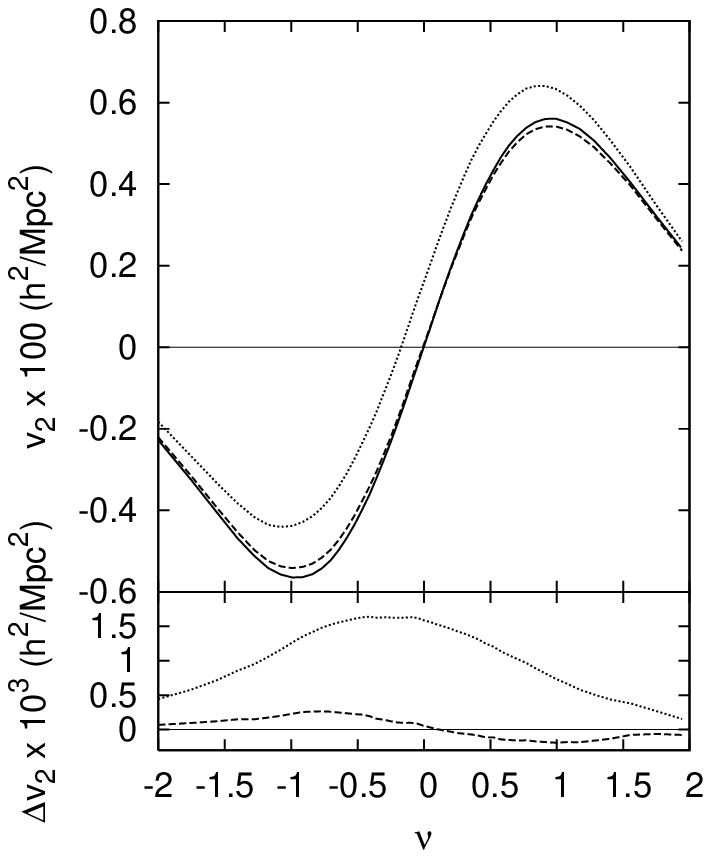}}
\resizebox{.32\textwidth}{!}{\includegraphics*{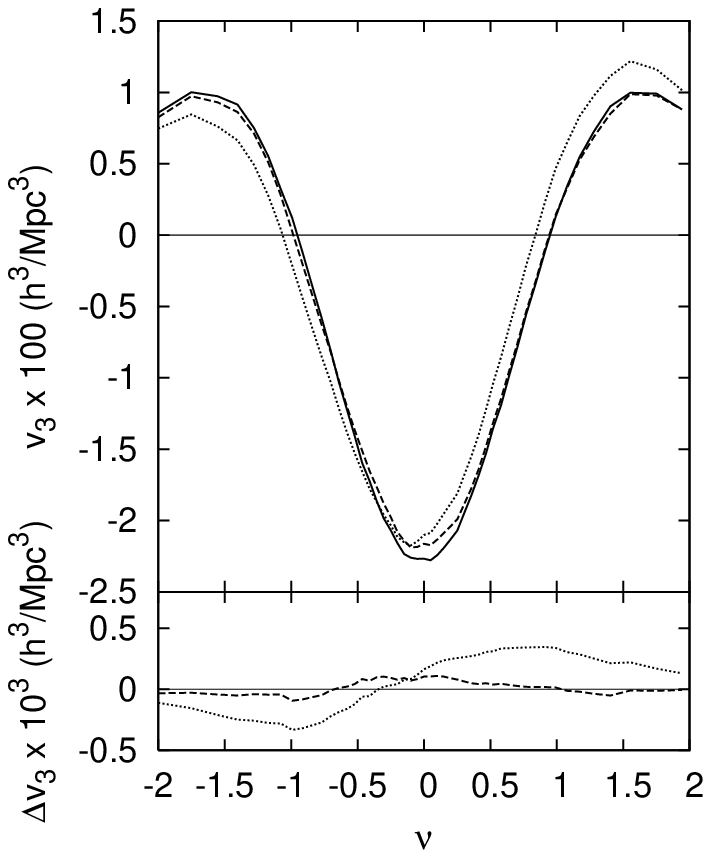}}
\caption{Demonstration of border corrections for complex
borders. The raw densities
of Minkowski functionals for the 2dFGRS NGP sample volume $252^3$,
cut from a  periodic $256^2\times64$ realization of a Gaussian
random field (dotted lines)
are shown together with the border-corrected estimates (dashed lines)
and the estimates for the original brick (full lines).
Upper panels show the densities, and smaller lower panels
show the differences between the densities for the sample volume
and those for the brick (for both the raw and corrected cases,
and the same line types are used as in the upper
panels).
The densities of the second MF $v_1$ are shown in the left panels, the
densities of the third MF $v_2$ -- in the middle panels, and the
densities of the fourth MF $v_3$ -- in the right panels.
\label{fig:Nmaskv123}}
\end{figure*}

    We tested these corrections with our original Gaussian realization,
masked at all faces. The correction for the second Minkowski
functional $v_1$ is practically perfect. The corrected version of
$v_2$ is also close to the original for all the argument range, and
only a little higher than the original. The higher the order of the
functional, the more difficult it is to correct for the borders, as
small errors from the lower orders accumulate. The discrepancy with
the corrected version of $v_3$ and the original estimates is the
largest amongst the three densities, but it balances well the
amplitudes of the maxima, and is only a little lower than the
original for low densities ($\nu\approx1.5$). There are practically
no differences from the original at the high-density end.

A note on the use of masks in practice: we ensure that there is at
least one-vertex thick mask layer around our data brick. This allows
us to assume periodic borders for the brick itself.
And there is also another way to use the mask, ignoring the vertices in
the mask, not building any elements from the vertices in the data
region to the mask vertices. Then we do not have to apply the
correction chain (\ref{corrchain}) and do not have to build the
basic sets in the mask region. The latter fact makes the
algorithm about twice as fast for the 2dF data (the data region
occupies only a fraction of the encompassing brick). We compared this
version with the border-corrected algorithm described above
(see Appendix~B)
and found that it gives slightly worse results for Gaussian
realizations, so we dropped it. The present algorithm is fast
enough, taking 12 seconds for a iso-level for a $256^3$ grid
on a laptop with the Intel Celeron 1500 MHz processor.

As the data masks are complex (see Fig.~\ref{fig:masks}), we should
test the border corrections for real masks, too.
Fig.~\ref{fig:Nmaskv123} shows the effect of border corrections as
used with the data mask for the 2dF NGC sample volume (see
section~\ref{sec:data}). As the corrections give densities of
functionals, we will show the densities from this point on. The
density field in this volume was generated by simulating a Gaussian
random field for a $256\times256\times64$ periodic brick, combining
these bricks to cover the sample extent, and masking this
realization with the Northern data mask. Smaller bricks were
combined because the spatial extent of the data was too large for
the available core memory to generate a single FFT brick to cover
it; as the brick is periodic, the realization remains Gaussian. We
show the raw densities of the Minkowski functionals as dashed lines,
the corrected versions with full lines, and the densities for the
original brick with dotted lines.

We see that the density $v_1$ of the $V_1$ functional is restored
well, apart from a slight deviation near $\nu=0$. The density $v_2$
is also corrected well, only its maximum amplitude is slightly
smaller than that for the original brick. The restoration is almost
perfect for $v_3$, with the same small amplitude problem than for
the two other densities. The fact that the restoration works so well
is really surprising -- first, our mask is extremely complex, and,
secondly, our realization of the Gaussian random field is certainly
not exactly homogeneous and isotropic inside the sample volume. We
generate realizations of random fields in this work by the FFT
technique; these fields are homogeneous, but isotropic only for
small scales, not for scales comparable to the brick size.

The importance of the good restoration of the Minkowski functional
means that when checking theoretical predictions, we can
directly compare observational results with the predictions,
and do not have to use costly Monte-Carlo simulations.

\subsection{Multiscale Minkowski Functionals}

A natural way to study cosmological fields is the multi-scale
approach, scale by scale. According to the present paradigm,
evolution of structure should proceed with different pace at
different scales. At smaller scales signatures of gravitational
dynamics should be seen, and traces of initial conditions could be
discovered at larger scales.

The matter density in the universe is formed by perturbations at all
scales. In the beginning they all grow at a similar rate, but soon
this rate becomes scale-dependent, the smaller the scale, the faster
it will go nonlinear and non-Gaussian. Thus it is interesting to
decompose the density (and gravitational potential, and velocity)
field into different scales and check their Gaussianity (and other
interesting characteristics).

Using the wavelet transform as it is described in Appendix~A, we
obtain a set of wavelet scales ${\cal W} = {w_0, ..., w_J,
c_{J+1}}$, and each scale $w_j(x,y,z)$ corresponds to the
convolution product of the observed galaxies with a wavelet function
$\psi_j$, where $\psi_j(x,y,z) = \psi(
\frac{x}{2^j},\frac{y}{2^j},\frac{z}{2^j})$ and $\psi$ is the
analyzing wavelet function described in Appendix~A. Now, we can
apply the MF calculation at each scale independently, and we get
four MF values per scale using Eq.~\ref{crofton}. The set $(V_{j,0},
V_{j,1}, V_{j,2}, V_{j,3})$ will denote the MF at scale $j$. Note
that in this framework, we do not have to convolve the data anymore
with a Gaussian kernel, avoiding the delicate choice of the size of
the kernel bandwidth.

\section{The data}
\label{sec:data} There exist two large-volume galaxy redshift
surveys at the moment, the Two Degree Field Galaxy Redshift Survey
(2dFGRS) \citep{2df} and the Sloan Digital Sky Survey (SDSS)
\citep{sdss}. The 2dFGRS is completed; although the SDSS is not, its
data volume has already surpassed that of the 2dFGRS. We shall use
in our paper the 2dFGRS dataset; it is easier to handle, and we make
use of the mock catalogues created to estimate the cosmic variance
of the data.

The galaxies of the 2dFGRS have been selected from an earlier
photometric APM survey \citep{maddox96} and its extensions. The
survey covers about 2000 deg$^2$ in the sky and consists of two
separate regions, one in the North Galactic Cap (NGC) and the other
in the South Galactic Cap (SGC), plus a number of small randomly
located fields; we do not use the latter. The total number of
galaxies in the survey is about 250,000. The depth of the survey is
determined by its limiting apparent magnitude, which was chosen to
be \mbox{$b_J=19.45$}. Caused by varying observing conditions,
however, this limit depends on the sky coordinates, varying almost
the full magnitude. Another cause of non-uniformity of the catalogue
is its spectroscopic incompleteness -- as the fibres used to direct
the light from a galaxy image in the focal plane to the spectrograph
have a finite size, a number of galaxies in close pairs were not
observed for redshifts. However, these corrections can be estimated;
the 2dFGRS team has made public the programs that calculate the
completeness factors and magnitude limit, given a line-of-sight
direction.

    The 2dFGRS survey, as all redshift surveys, is magnitude-limited.
This means that the density of observed galaxies decreases with
distance; at large distances only intrinsically brighter galaxies
can be seen. For certain statistical studies (luminosity functions,
correlation functions, power spectra) this decrease can be corrected
for. For texture studies there are yet no appropriate correction
methods, and maybe, these do not exist, as the scales of the details
that can be resolved are inevitably different, and small-scale
information is certainly lost at large distances. The usual approach
is to use volume-limited subsamples extracted from the survey. In
order to create such a sample, one chooses absolute magnitude
limits, and retains only the galaxies with absolute magnitudes
between these limits. This discards most of the data, but assures
that the spatial resolution is the same throughout the survey volume
(taking also account of the possible luminosity evolution and
K-correction).

    The 2dFGRS team has created such catalogues and used them to
study higher-order correlation functions
\citep{norberg02,croton1,croton2}. They kindly made these catalogues
available to us, and these constitute our main data. These
catalogues span one magnitude each, from $M= -17+5\log10(h)$ until
$M= -21+5\log10(h)$. The catalogues for least bright galaxies span
small spatial volumes, and those for the brightest galaxies are
sparsely populated. Thus we chose for our work the catalogue
spanning the magnitude range $-20\le M-5\log10(h)\le-19$, this is
the most informative. This has been also the conclusion of
\citet{croton2}. We shall call this sample 2dF19. This sample, as
all 2dF volume-limited samples, consists of two spatially distinct
subsamples, one in the North Galactic Cap region (2df19N) and
another in the South Galactic Cap region (2df19S). The sample lies
between 61.1~Mpc/$h$ and 375.6~Mpc/$h$; the general features of the
subsamples are listed in Table~\ref{tab:cats}.

\begin{table*}
\centering \caption{The 2dF volume-limited catalogues used.
\label{tab:cats}}
\medskip

\begin{tabular}{|c|r|rr|rr|r|r|}\hline
sample&galaxies&\multicolumn{2}{c|}{ra limits (deg)}&
\multicolumn{2}{c|}{dec limits}& \multicolumn{1}{c|}{Vol ($10^6$}&
\multicolumn{1}{c|}{dmean}\\
&&\multicolumn{2}{c|}{(deg)}& \multicolumn{2}{c|}{(deg)}&
\multicolumn{1}{c|}{Mpc$^3h^{-3}$)}&
\multicolumn{1}{c|}{(Mpc/$h$)}\\
\hline
2dF19N&19080&147.0&223.0&$-$6.4&2.6&2.75&5.24\\
2dF19S&25633&$-$35.5&55.2&$-$37.6&$-$22.4&4.43&5.57\\
\hline
\end{tabular}
\end{table*}

In a previous paper on the morphology of the 2dFGRS \citep{mart05}
we extracted bricks from the data to avoid the influence of border
effects. This forced us to use only a fraction of the volume-limited
samples. This time we tried to use all the available data, and
succeeded with that.
\begin{figure}
\centering
\resizebox{.48\textwidth}{!}{\includegraphics*{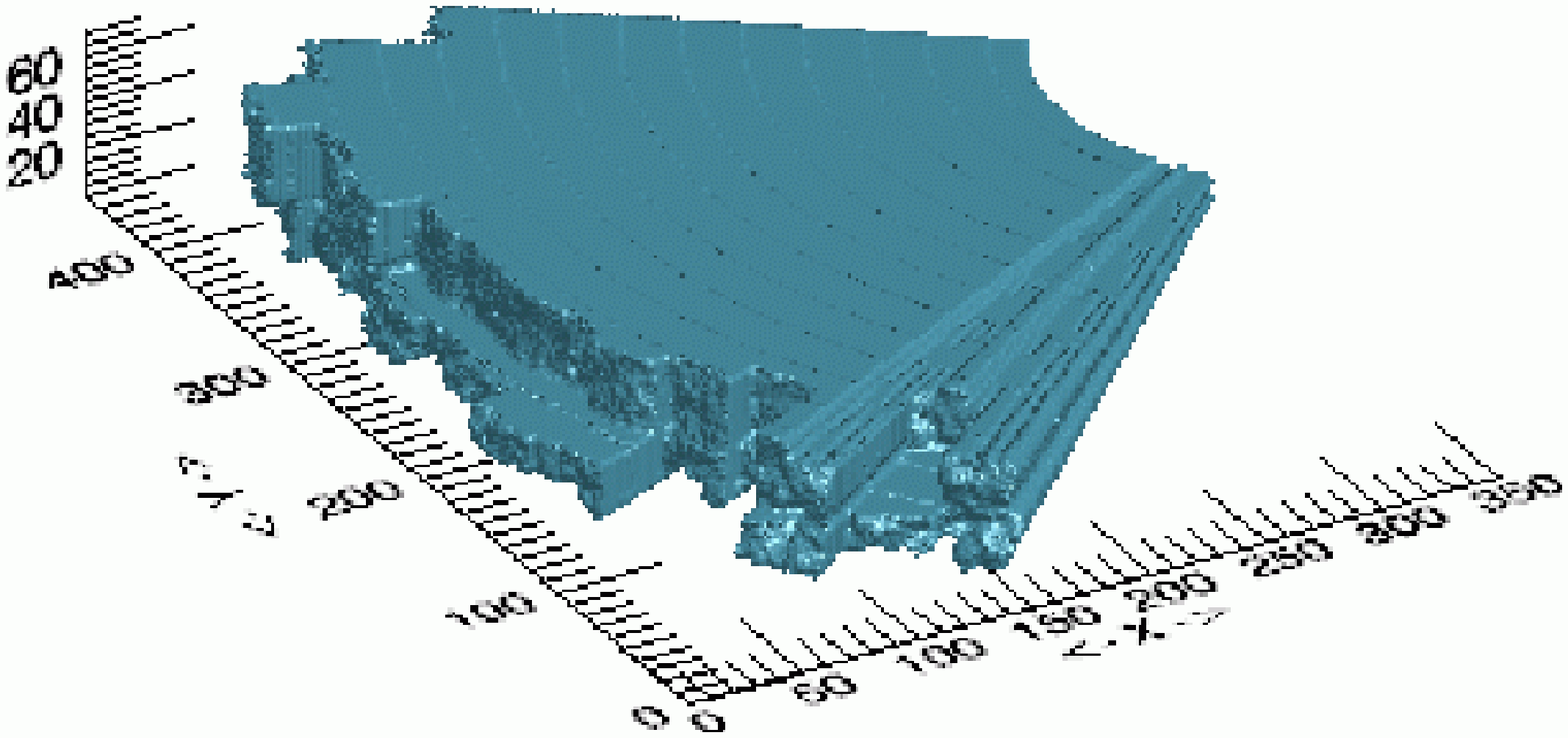}}\\
\resizebox{.48\textwidth}{!}{\includegraphics*{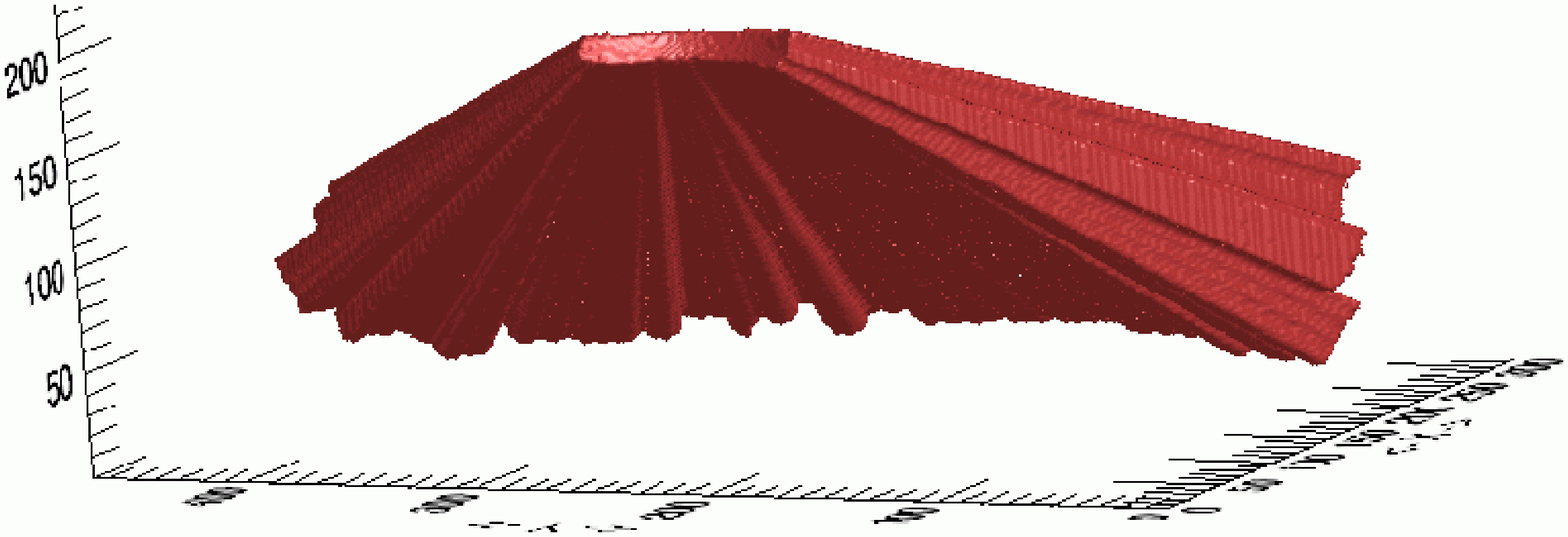}}\\
\caption{Survey masks for the 2dFGRS volume-limited sample 2dF19.
Upper panel -- the Northern subsample, lower panel -- the Southern
subsample. Spatial orientations are chosen to better visualise the
volumes.\label{fig:masks}}
\end{figure}

The 2dFGRS catalogue is composed of measurements in a large number
of circular patches in the sky, and its footprint in the sky is
relatively complex (see the survey web-page
\texttt{http://www.mso.anu.au/2dFGRS}). Furthermore, due to the
variations in the final magnitude limit used, the catalogue depth is
also a function of the direction. In order to use all the data for
wavelet and morphological analysis, we had to create a spatial mask,
separating the sample volume from regions outside. We did it by
creating first a spatial grid for a brick that surrounded the
observed catalogue volume, calculated the sky coordinates for all
vertices of the brick, and used then the software provided by the
2dFGRS team to find the correction factors for these directions. The
completeness factor told us if the direction was inside the sample
footpath. If it was, we found the apparent magnitude of the
brightest galaxies (with $M=M_{\mbox{\scriptsize min}}$) of our
sample for that distance and checked, if it was lower than the
sample limit. If it was, the grid point was included in the mask.
For the last comparison we had to change from the comoving grid
distance to the luminosity distance. We assumed the $\Omega_M=0.3$,
$\Omega_\Lambda=0.7$ cosmology for that and interpolated a tabulated
relation between the two distances. As explained in
\citet{norberg02}, the 2dF volume-limited samples were built using
the $k+e$-correction as dependent on the spectral type of a galaxy.
We do not have such a quantity for the mask, so we tuned a little
the bright absolute magnitude limit, checking that the mask should
extend as far as the galaxy sample. For that, we had to increase the
effective bright absolute magnitude limit by $\Delta M= 0.25$. The
nearby regions of the mask were cut off at the nearest distance
limit for the observational sample.

The original survey mask in the sky includes holes around bright
stars; these holes generate narrow tunnels through our spatial mask.
As the masks have a complex geometry, we did not want to add
discreteness effects due to resolving such tunnels. We filled them
in by counting the number of neighbours in a $3^3$ cube around
non-mask points and, if the neighbour number was larger than a
chosen limit, assigning the points to the mask. We chose the
required number $n$ of neighbours to avoid filling in at flat mask
borders ($n=9$ is enough for that) and iterated the procedure until
the tunnels disappeared. This was determined by visual checks (using
the 'ds9' fits file viewer \citep{ds9}).

    We show the 3-D views of our masks in Fig.~\ref{fig:masks}. The
mask volumes are, in general, relatively thin curved slices with
heavily corrugated outer walls. These corrugations are caused by the
unobserved survey fields. Also, the outer edges of the mask are
uneven, due to the variations in the survey magnitude limit. One 3-D
view does not give a good impression of the mask; the slices we
shall show below will complement these.

As mentioned in Appendix~A, in order to apply the wavelet
convolution cascade, the initial density on the grid should be
extirpolated by the chosen scaling function. We chose our initial
grid step as 1~Mpc/$h$, and used the $B^{(3)}_3$ kernel for
extirpolation. In order to have a better scale coverage, we repeated
the analysis, using the grid step $\sqrt2$~Mpc/$h$. The smoothing
scale (the spatial extent of the kernel) is 4 grid units, the
smoothing radius corresponds to 2 units. When calculating the
densities, we used the spectroscopic completeness corrections
$c_{\mbox{i\scriptsize sp}}$, included in the 2dFGRS volume-limited
catalogues, and weighted the galaxies by the factor
$w=1/c_{\mbox{i\scriptsize sp}}$. Most of the weights are close to
unity, but a few of them are large. In order not to 'overweight'
these galaxies, we fixed the maximum weight level as $w=2$. The same
procedure was chosen by \citet{croton1}.


\section{Gaussian fields}
\label{sec:gauss}

The filters used to perform wavelet expansion
are linear, and thus should keep the morphological
structure of Gaussian fields; the Minkowski functionals
should be Gaussian for any wavelet order.
This is certainly true for periodic densities,
but for densities restricted to finite volumes
the boundary conditions can introduce correlations.
The most popular boundary condition -- reflection at
the boundary -- will keep the density field mostly
Gaussian for brick masks. Our adopted zero boundary condition
will certainly work destroying Gaussianity, as the
random field which is zero outside a given
volume and has finite values inside is certainly not Gaussian.

\begin{figure*}
\centering
\resizebox{0.32\textwidth}{!}{\includegraphics*{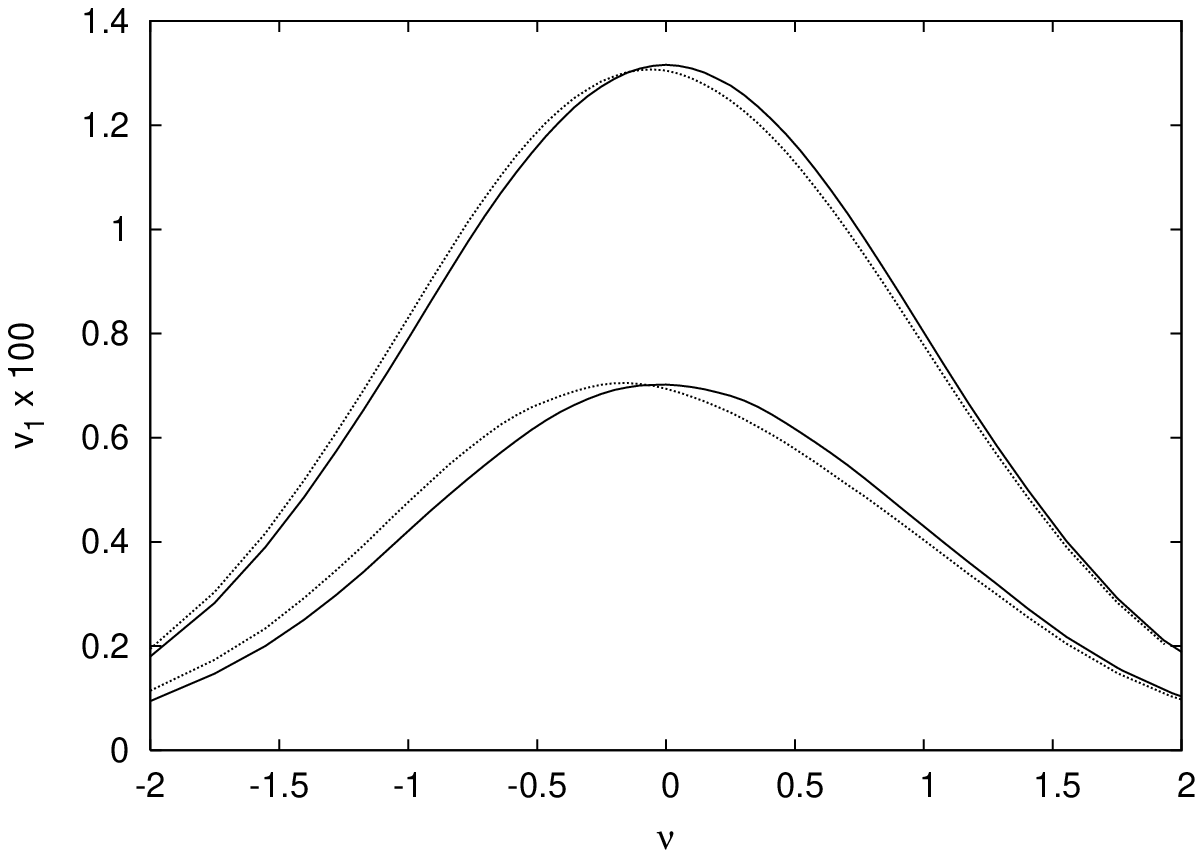}}
\resizebox{0.32\textwidth}{!}{\includegraphics*{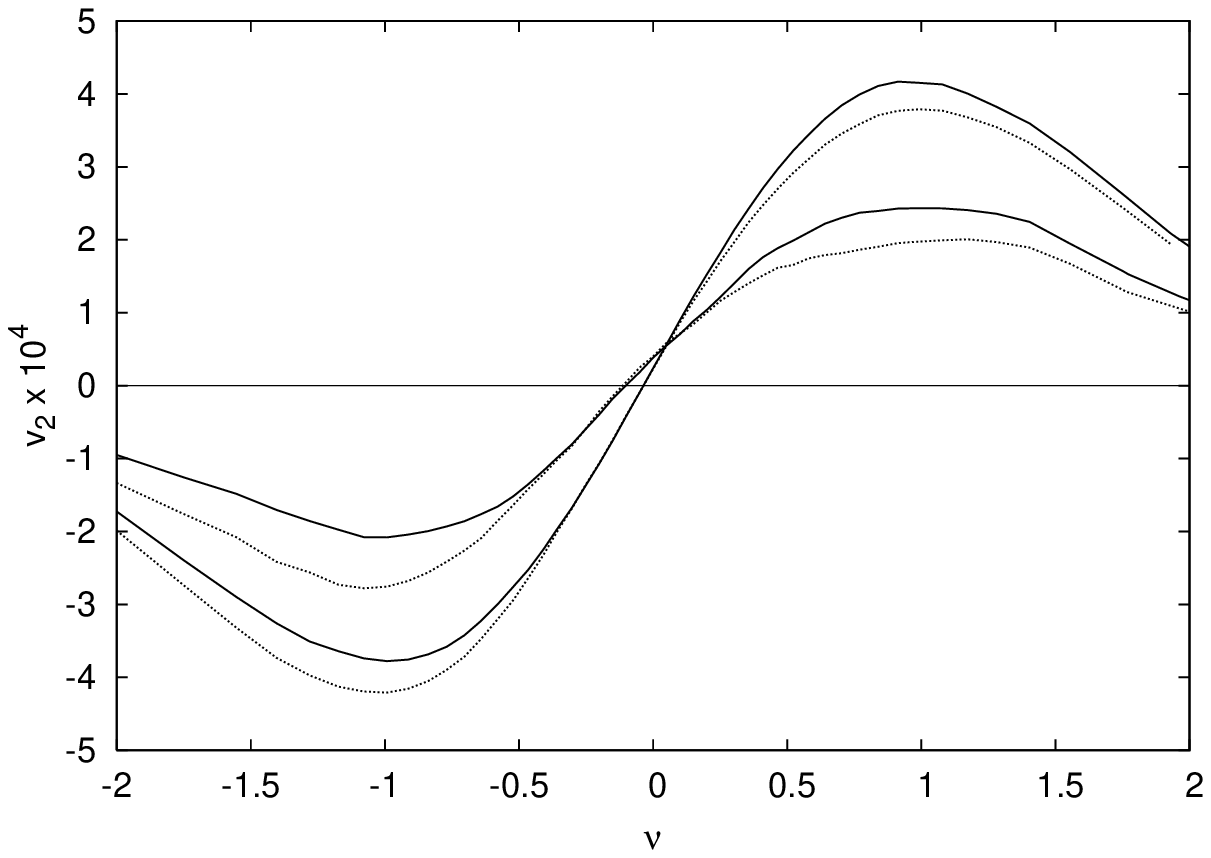}}
\resizebox{0.32\textwidth}{!}{\includegraphics*{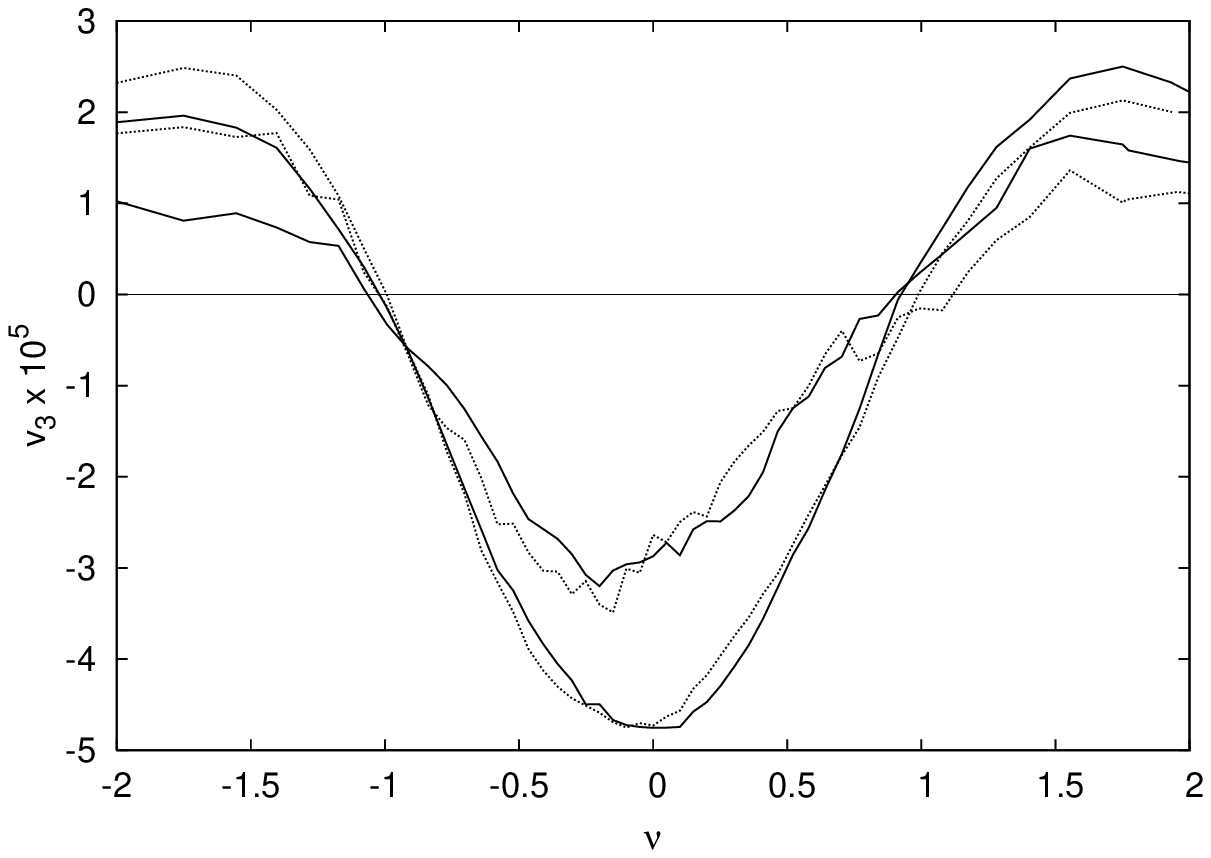}}
\caption{Densities of the Minkowski functionals (from the second
at the left until the fourth at the right)
for the same Gaussian realization for
the wavelet orders 3 and 4 (lower wavelet orders give higher
amplitudes). The case for the wavelets generated using the mirror
boundary conditions is shown by full lines, for zero boundary
conditions dotted lines are used. The amplitudes of the functionals
for the wavelet order 4 are rescaled by 2 for $v_2$ and by 4 for $v_3$
to show more details.
\label{fig:mfcomp}}
\end{figure*}

\begin{figure*}
\centering
\fbox{\resizebox{0.25\textwidth}{!}{\includegraphics*{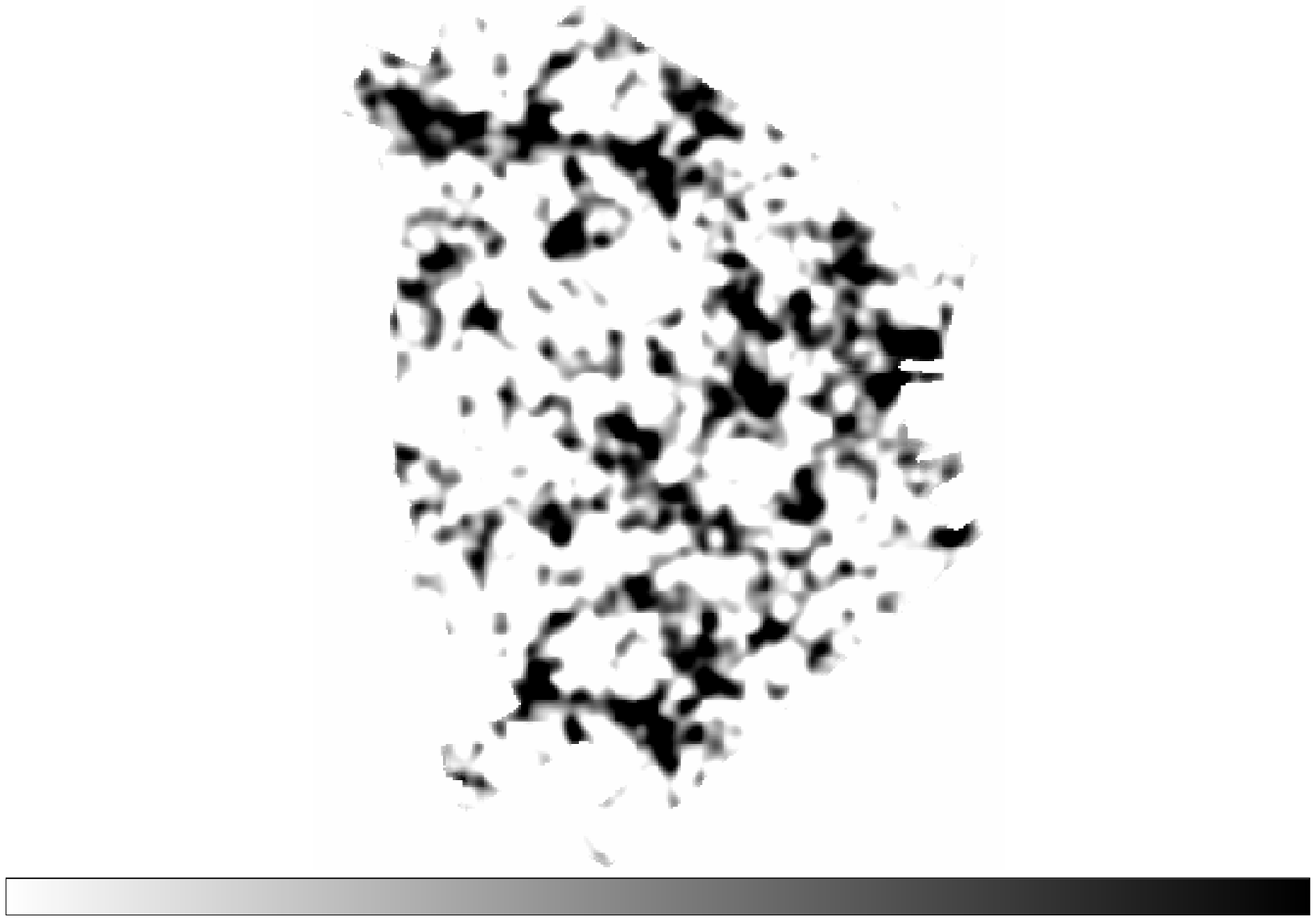}}}~
\fbox{\resizebox{0.25\textwidth}{!}{\includegraphics*{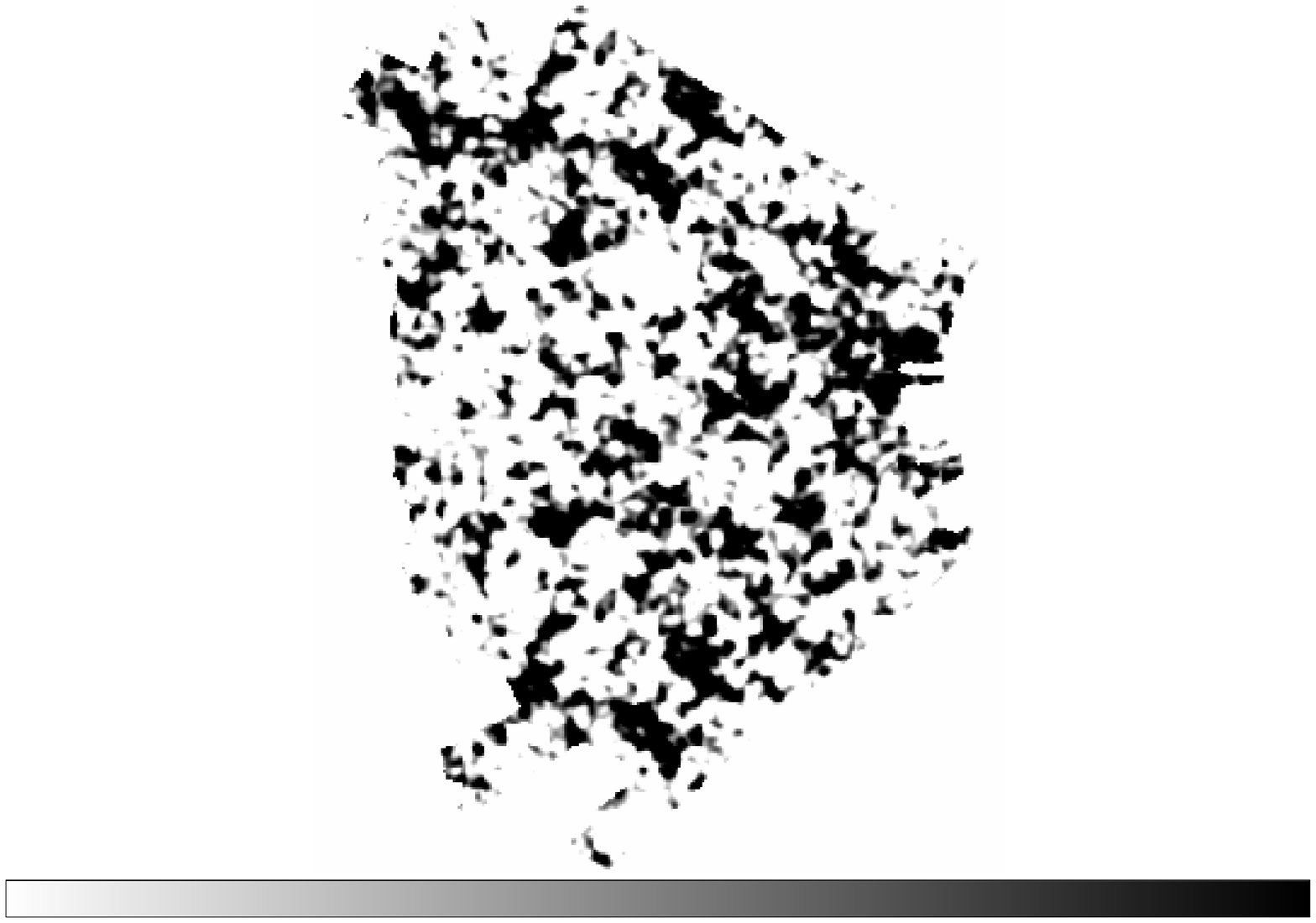}}}~
\fbox{\resizebox{0.25\textwidth}{!}{\includegraphics*{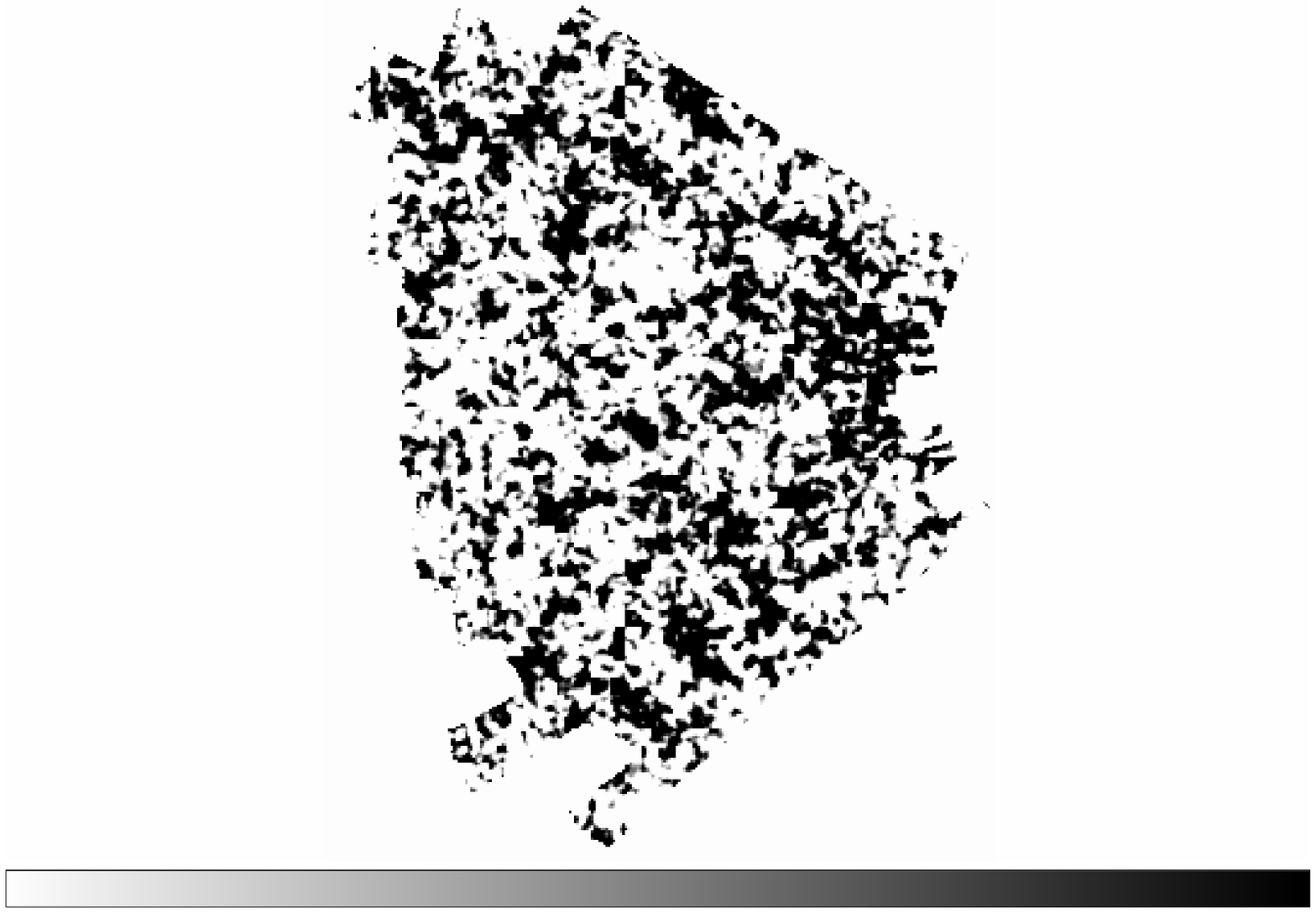}}}\\[4pt]
\fbox{\resizebox{0.25\textwidth}{!}{\includegraphics*{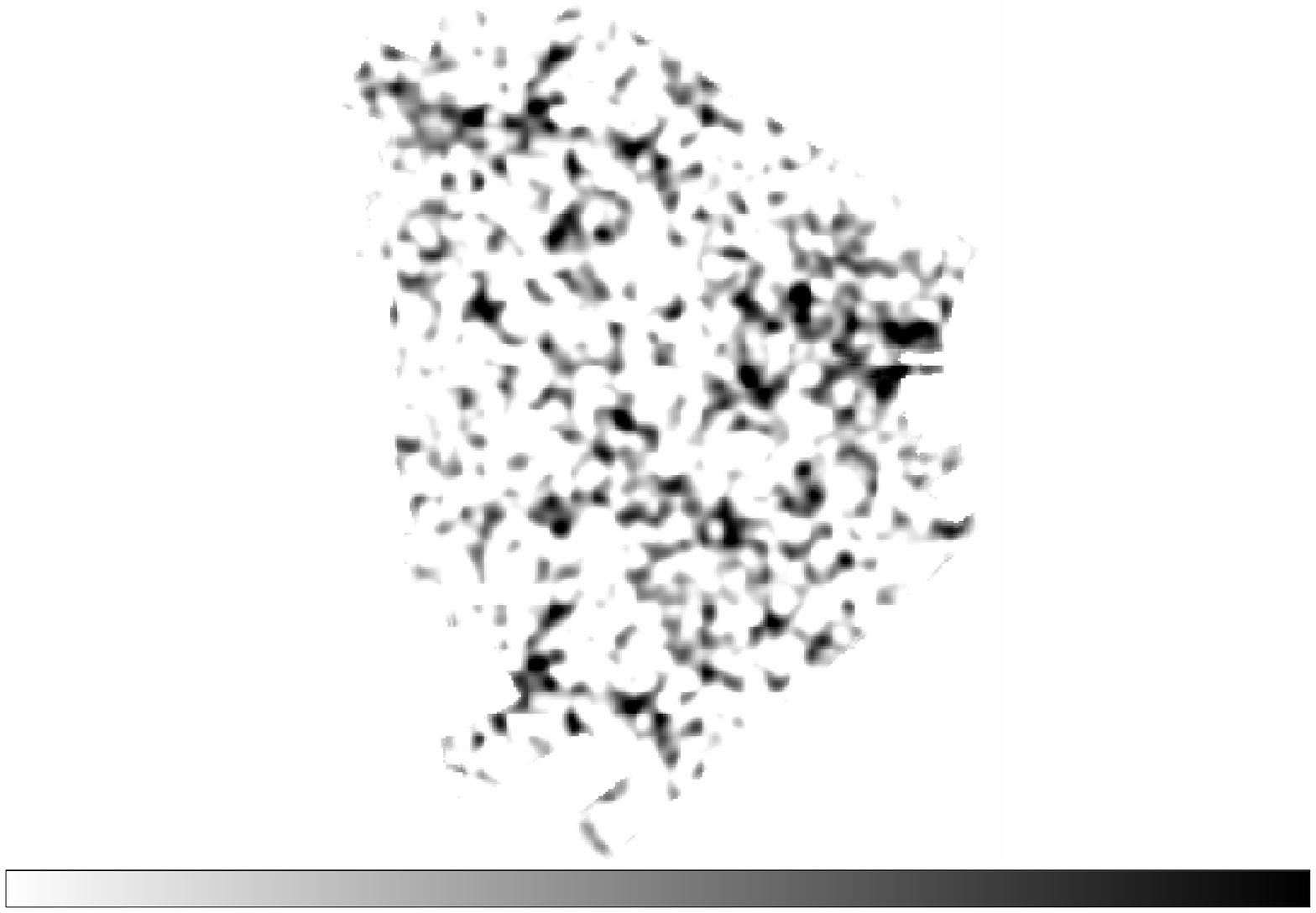}}}~
\fbox{\resizebox{0.25\textwidth}{!}{\includegraphics*{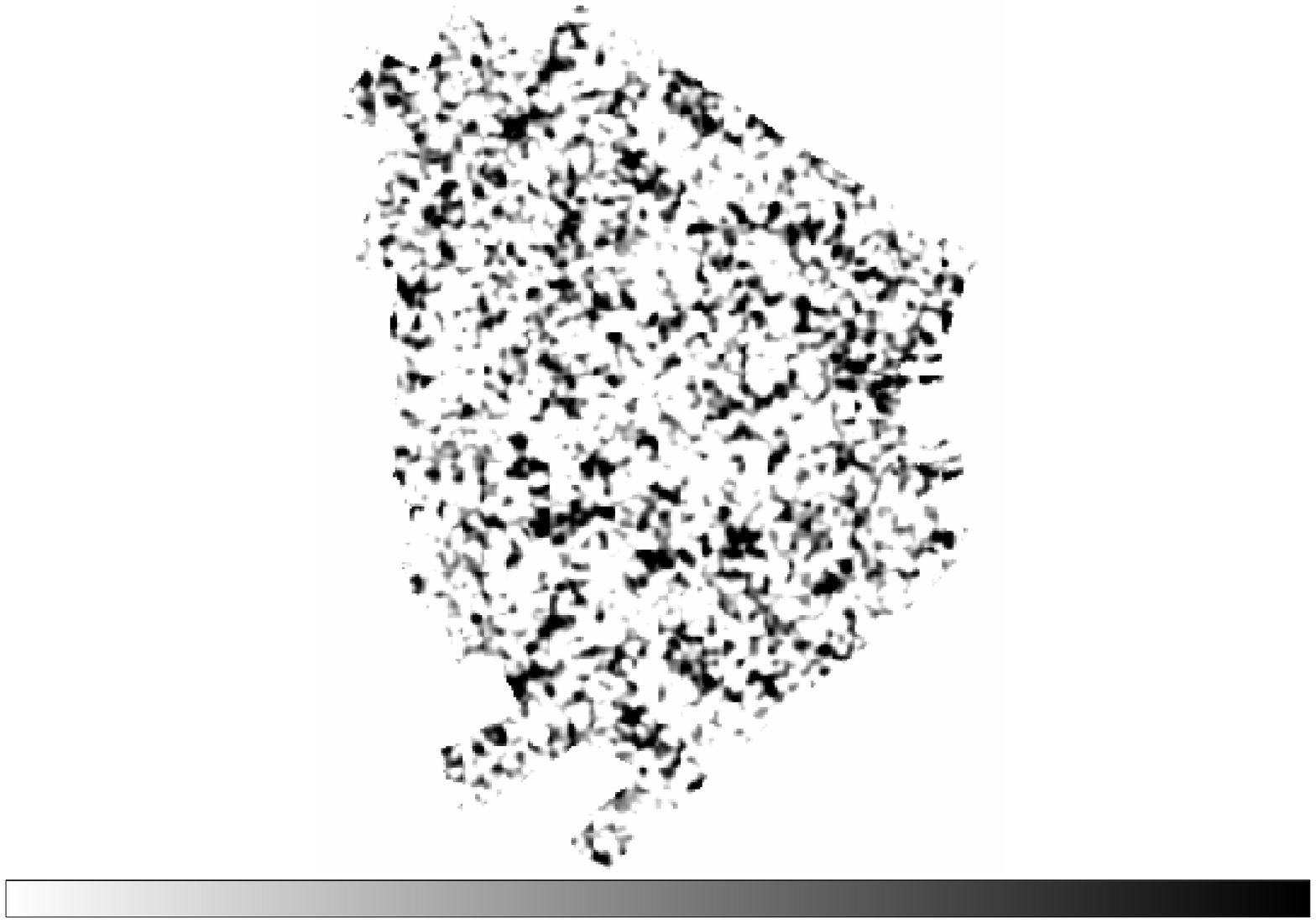}}}~
\fbox{\resizebox{0.25\textwidth}{!}{\includegraphics*{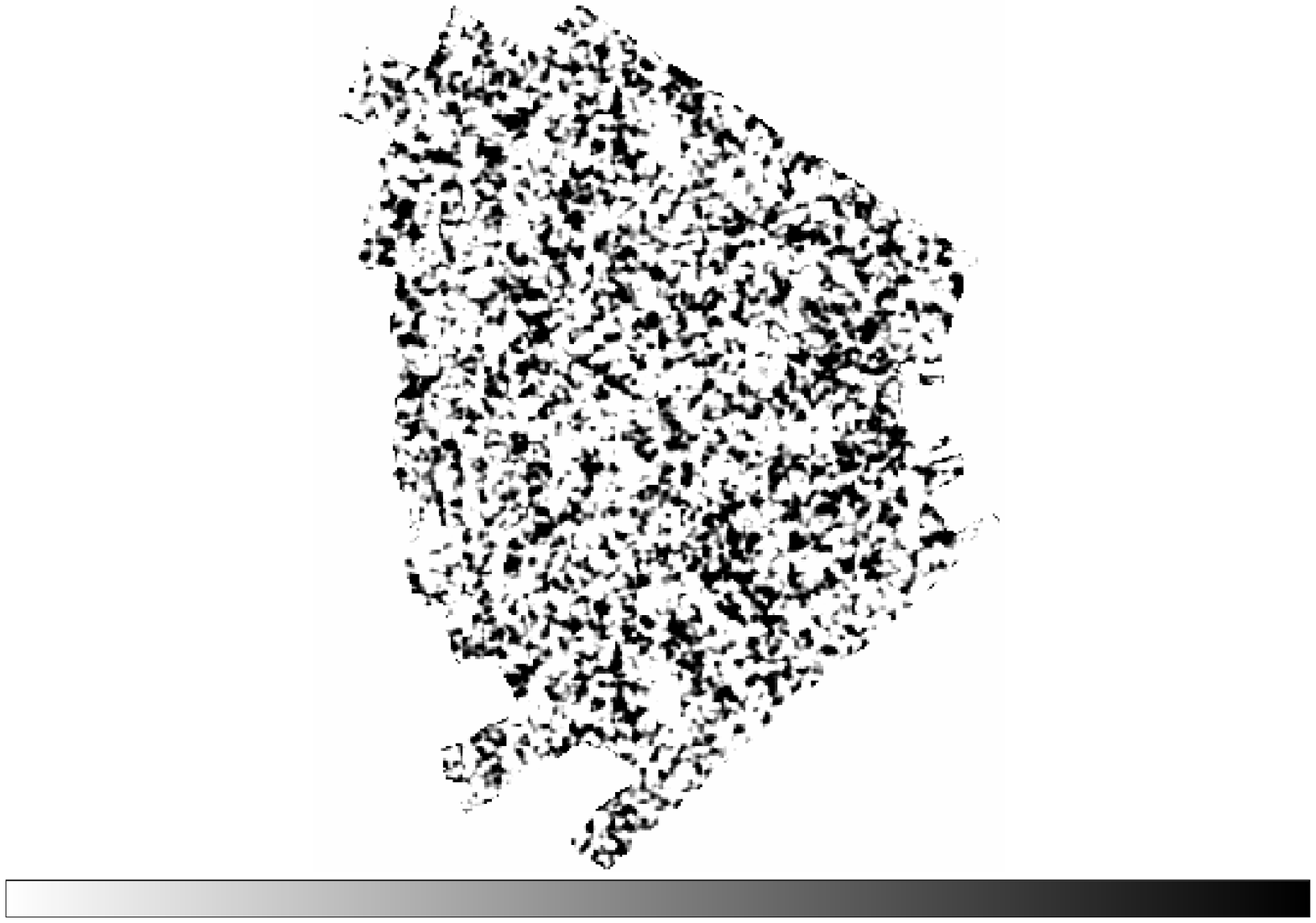}}}\\
\caption{Multi-scale decomposition of a realization of a Gaussian
random field in the 2df19N volume mask, for the $z=34$~Mpc/$h$
slice (the data and the first orders). The upper panel column shows
the scaling orders, with the original density at the right.
The lower panel shows the wavelet orders, with the lowest
order at the right.
\label{fig:Ngsbunch1}}
\end{figure*}

\begin{figure*}
\centering
\newlength{\pushlen}
\setlength{\pushlen}{0.25\textwidth}
\addtolength{\pushlen}{2\fboxsep}
\addtolength{\pushlen}{2\fboxrule}
\fbox{\resizebox{0.25\textwidth}{!}{\includegraphics*{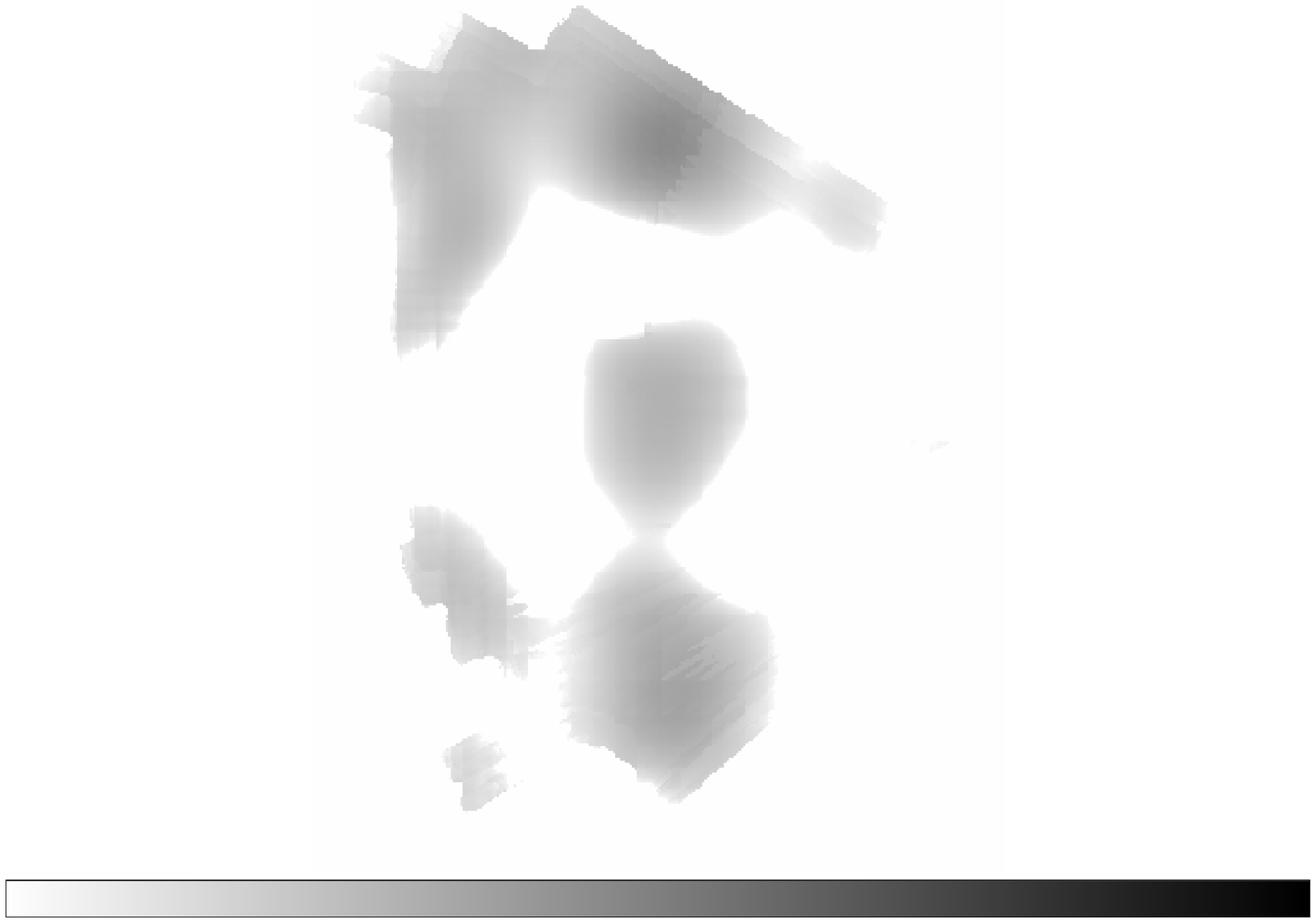}}}~
\fbox{\resizebox{0.25\textwidth}{!}{\includegraphics*{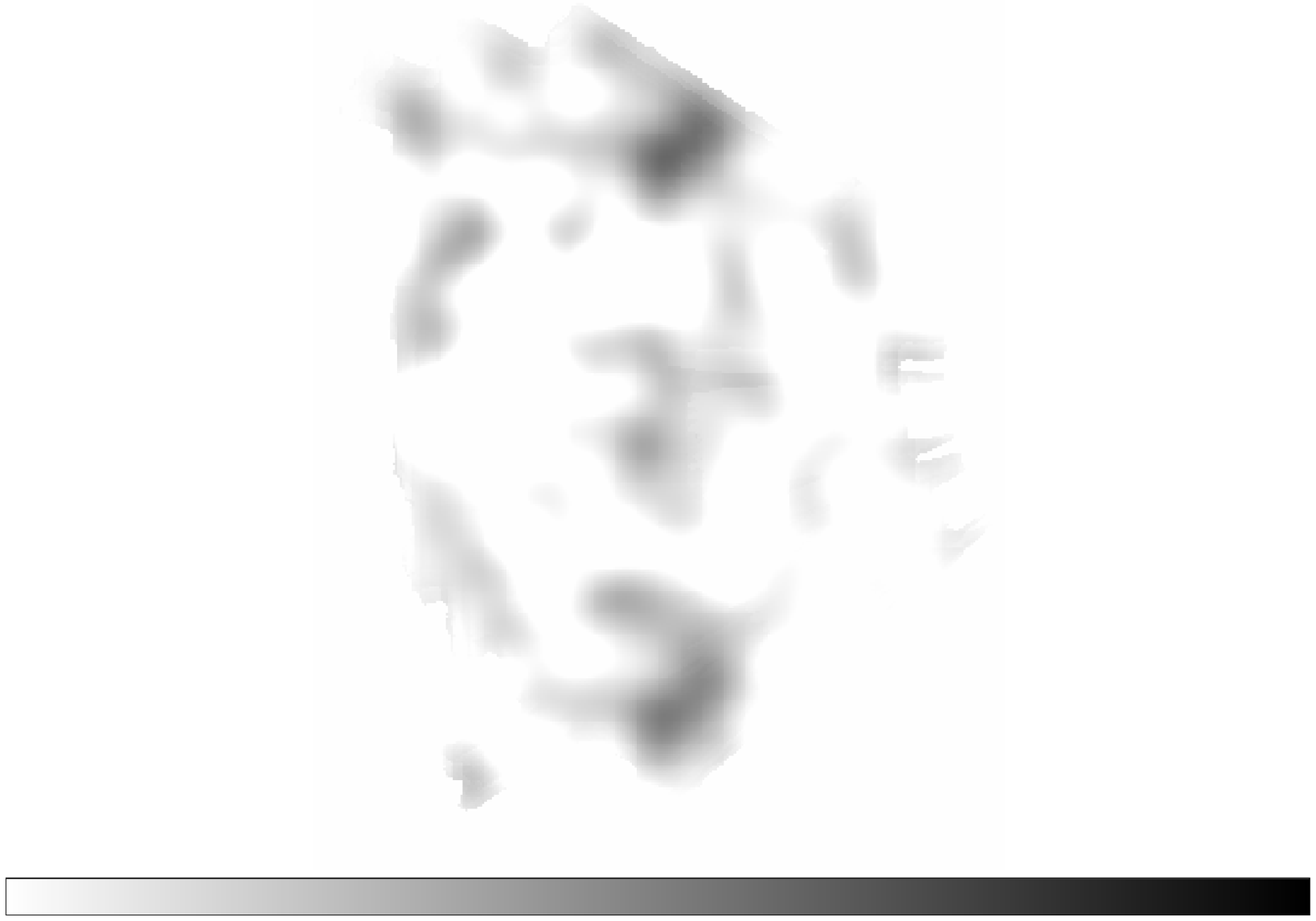}}}~
\fbox{\resizebox{0.25\textwidth}{!}{\includegraphics*{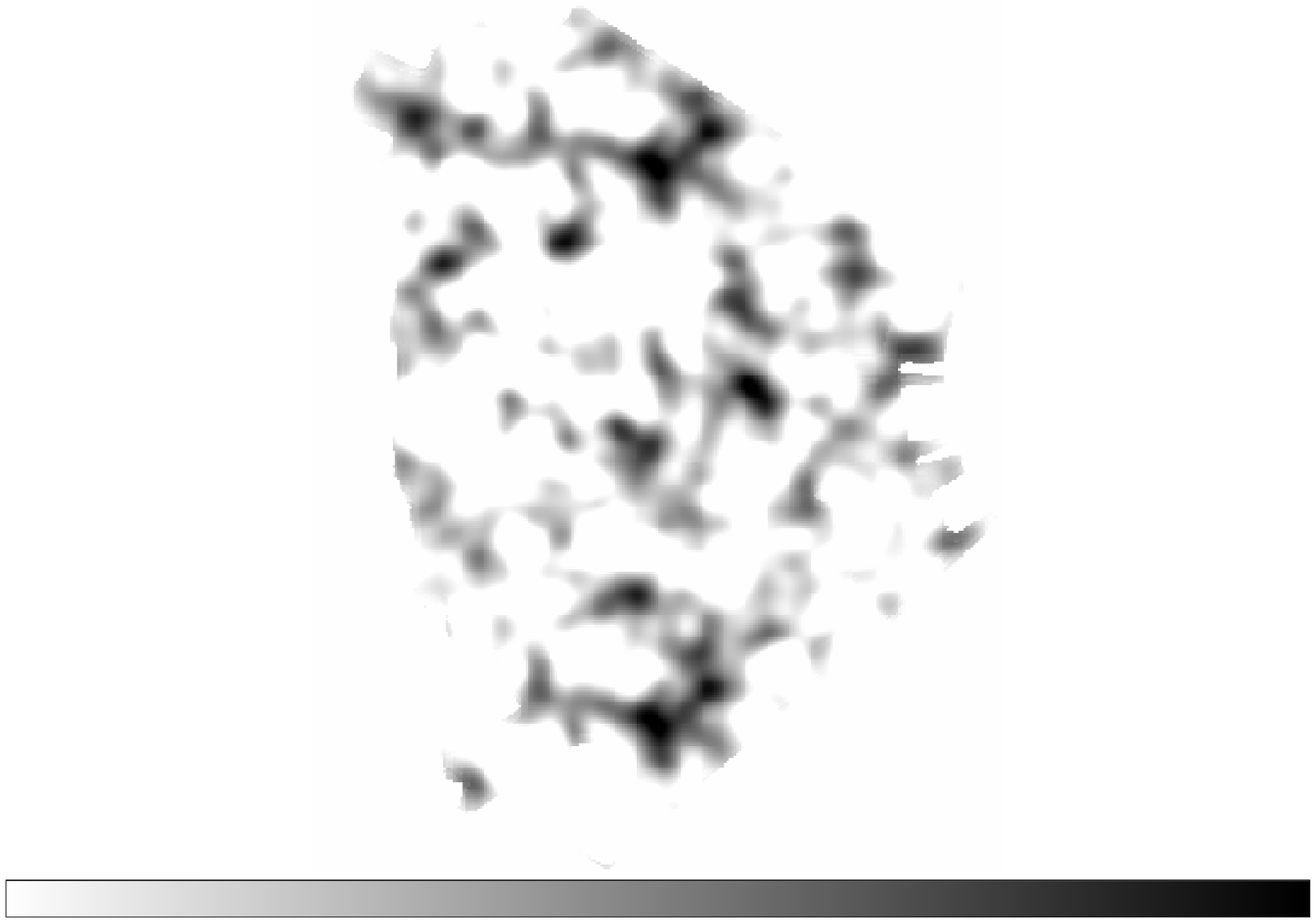}}}\\[4pt]
\parbox{\pushlen}{\hfill}~
\fbox{\resizebox{0.25\textwidth}{!}{\includegraphics*{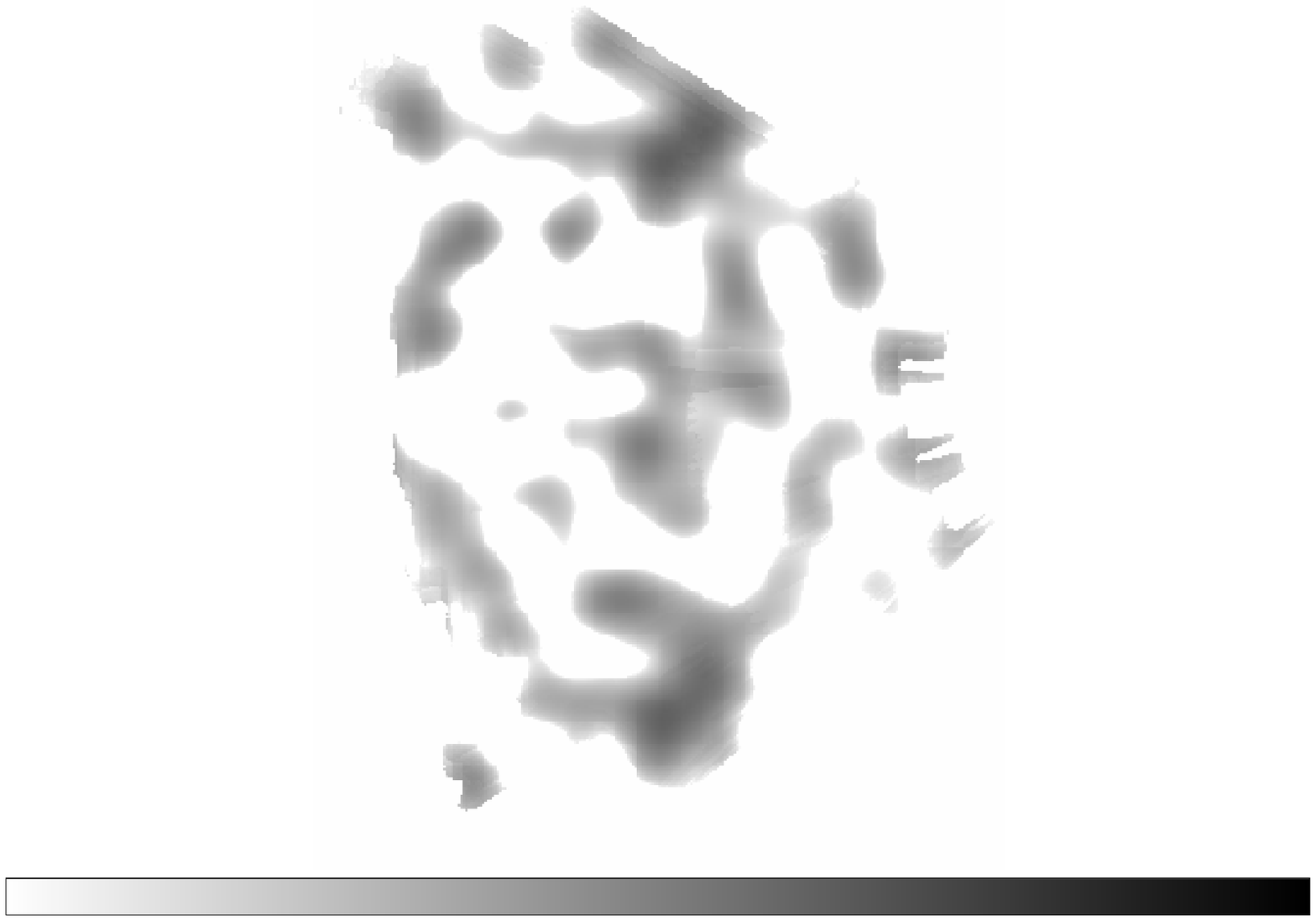}}}~
\fbox{\resizebox{0.25\textwidth}{!}{\includegraphics*{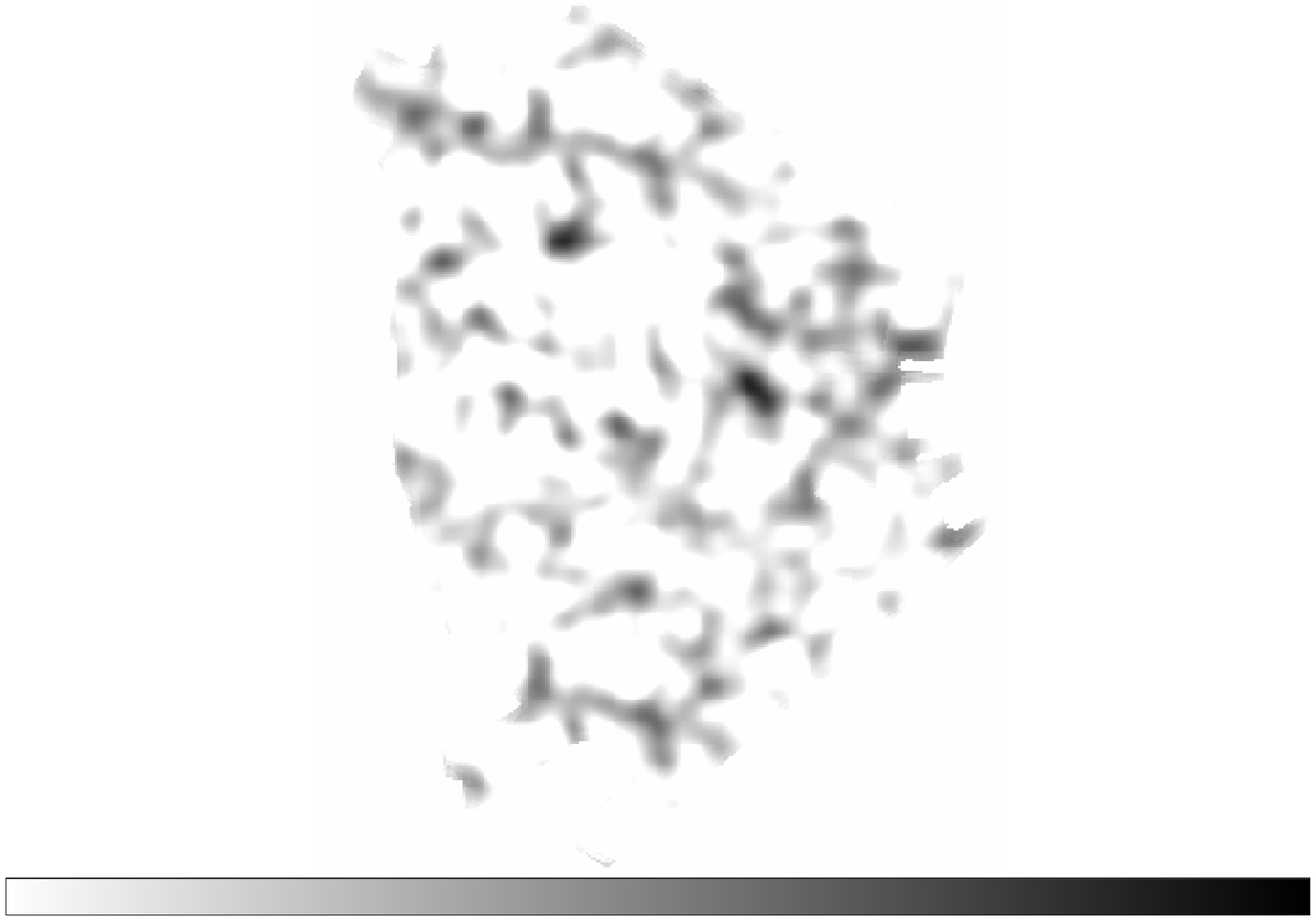}}}\\
\caption{Multi-scale decomposition of of a Gaussian
random field in the 2df19N volume mask (continued).
Upper panel -- scaling orders, lower panel -- wavelet orders,
the highest orders at the left.
The last scaling solution shows
already strong effects of the boundary conditions.
\label{fig:Ngsbunch2}}
\end{figure*}

We compare the effect of the mirror
and zero boundary conditions in Fig.~\ref{fig:mfcomp},
for high wavelet orders that are yet not dominated by noise,
calculated for a Gaussian realization in a $256^3$ cube
with the $\sigma=3$ Gaussian smoothing described above, and
masked at all borders by a layer two vertices thick
(thus the effective volume of the data cube is $252^3$)

Fig.~\ref{fig:mfcomp} shows that the second Minkowski functional
is certainly better restored by the wavelets obtained by using
mirror boundary conditions.
These conditions lead to the functionals that
are symmetric about $\nu=0$, as it should be, while the functionals
obtained by applying zero boundary conditions display a shift towards smaller
values of $\nu$.
However, although our brick data
should prefer mirror boundary conditions, it is not easy to say which
boundary conditions give better estimates for the remaining two functionals.
Both cases have comparable errors (look at the amplitudes at extrema,
which should be equal). Hence, other considerations have to be used.
As our data mask is complex, with corrugated planes and sharp corners,
the mirror boundary conditions will amplify the corner densities and
propagate them inside, while the zero boundary conditions will
gradually remove the influence of the boundary data. Thus we have
selected zero boundary conditions for the present study; these are
probably natural for all observational samples.

\begin{figure}
\centering
\resizebox{0.48\textwidth}{!}{\includegraphics*{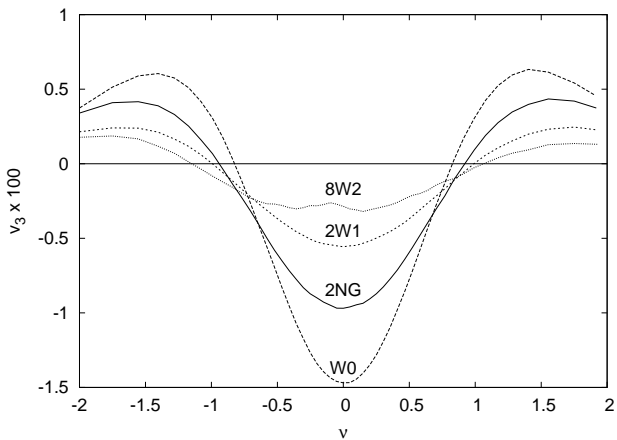}}
\caption{The density $v_3$ of the fourth MF
for the wavelet decomposition of a realization of a Gaussian
random field in the 2dfN19 mask. The legends in the figure show
the wavelet order and the scaling factor (2W1 denotes the
functional for the wavelet order 1, multiplied by 2). The
legend NG denotes the original realization.
\label{fig:mf3wng}}
\end{figure}

\begin{figure*}
\centering
\fbox{\resizebox{0.25\textwidth}{!}{\includegraphics*{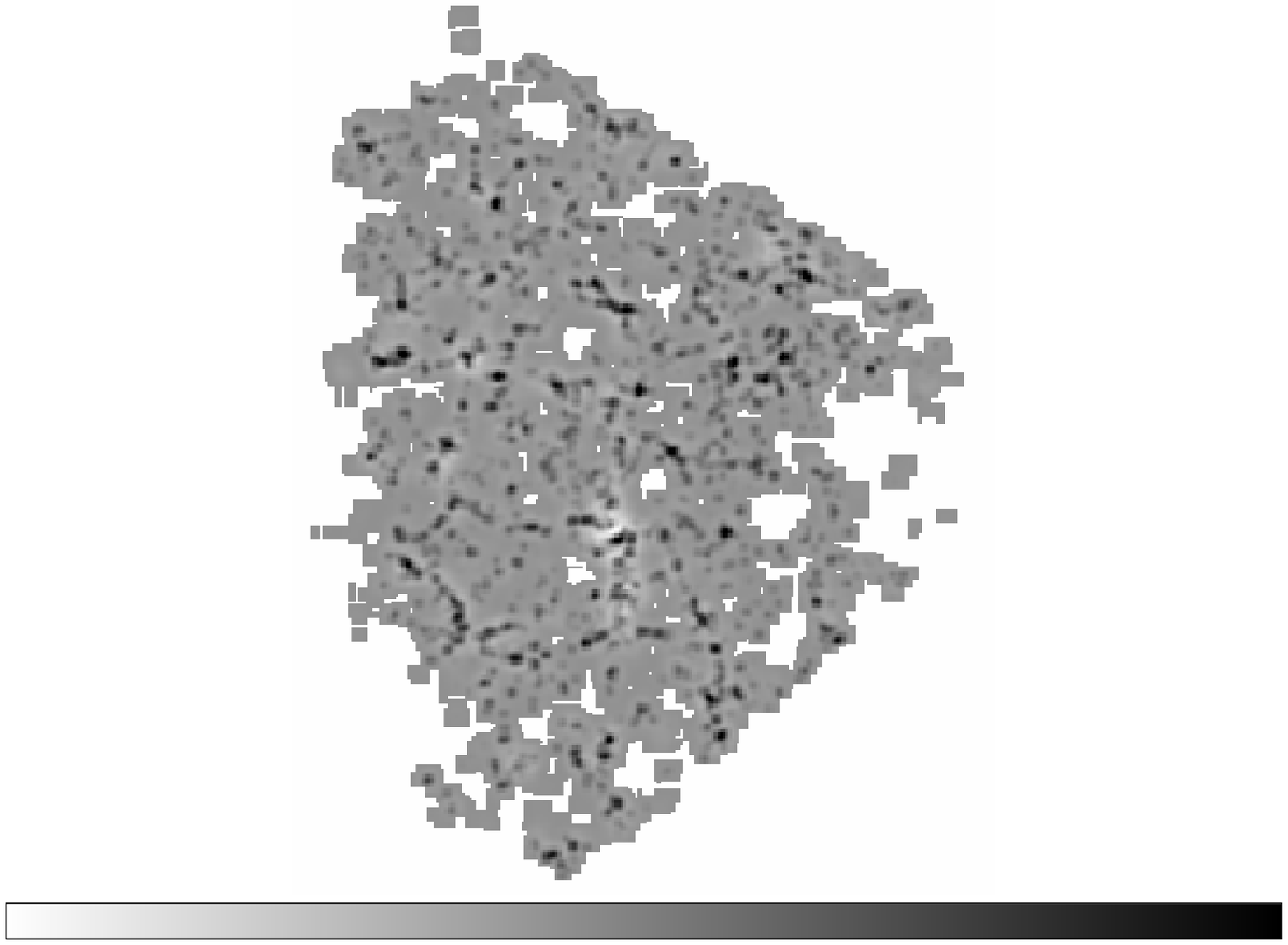}}}~
\fbox{\resizebox{0.25\textwidth}{!}{\includegraphics*{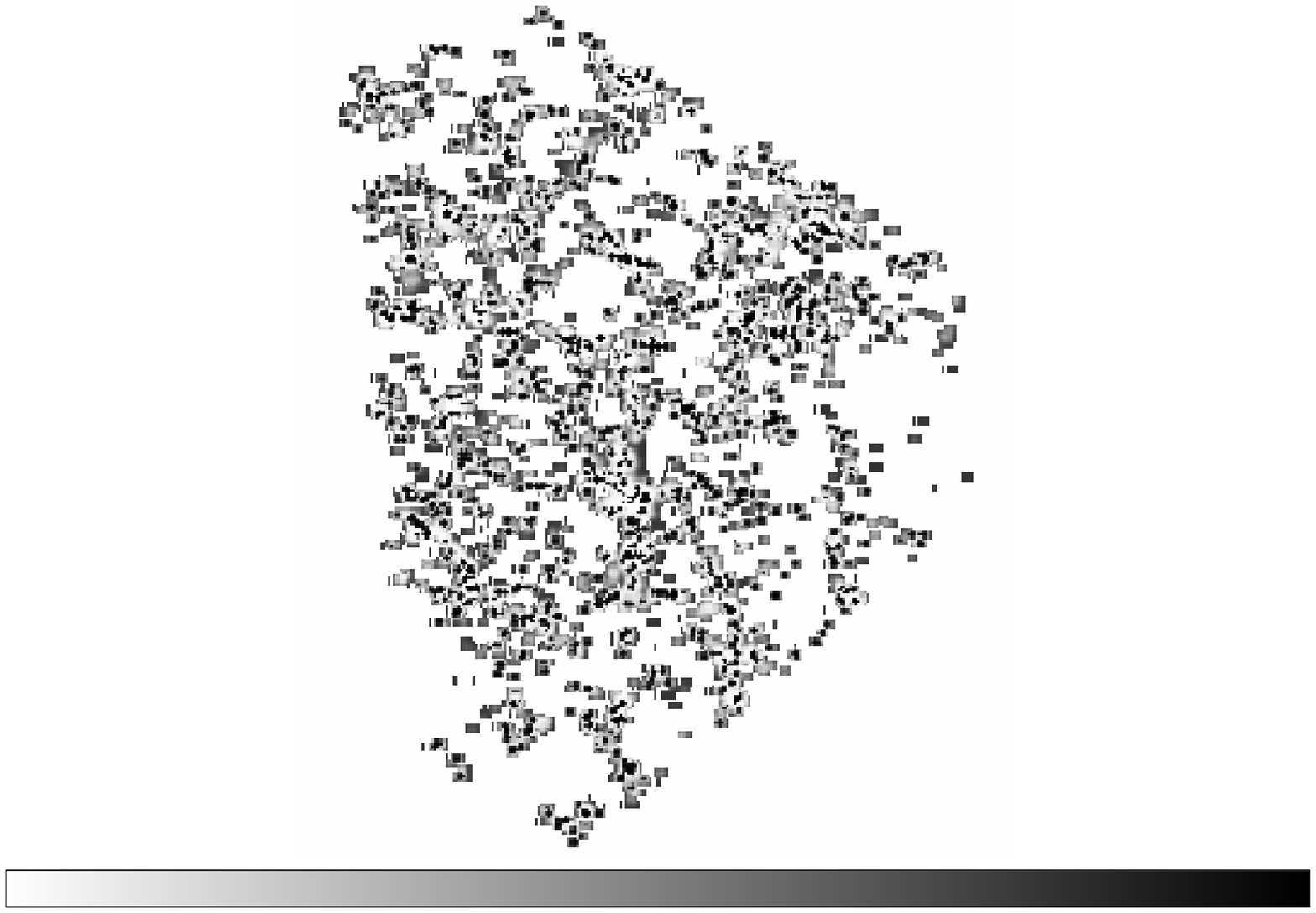}}}~
\fbox{\resizebox{0.25\textwidth}{!}{\includegraphics*{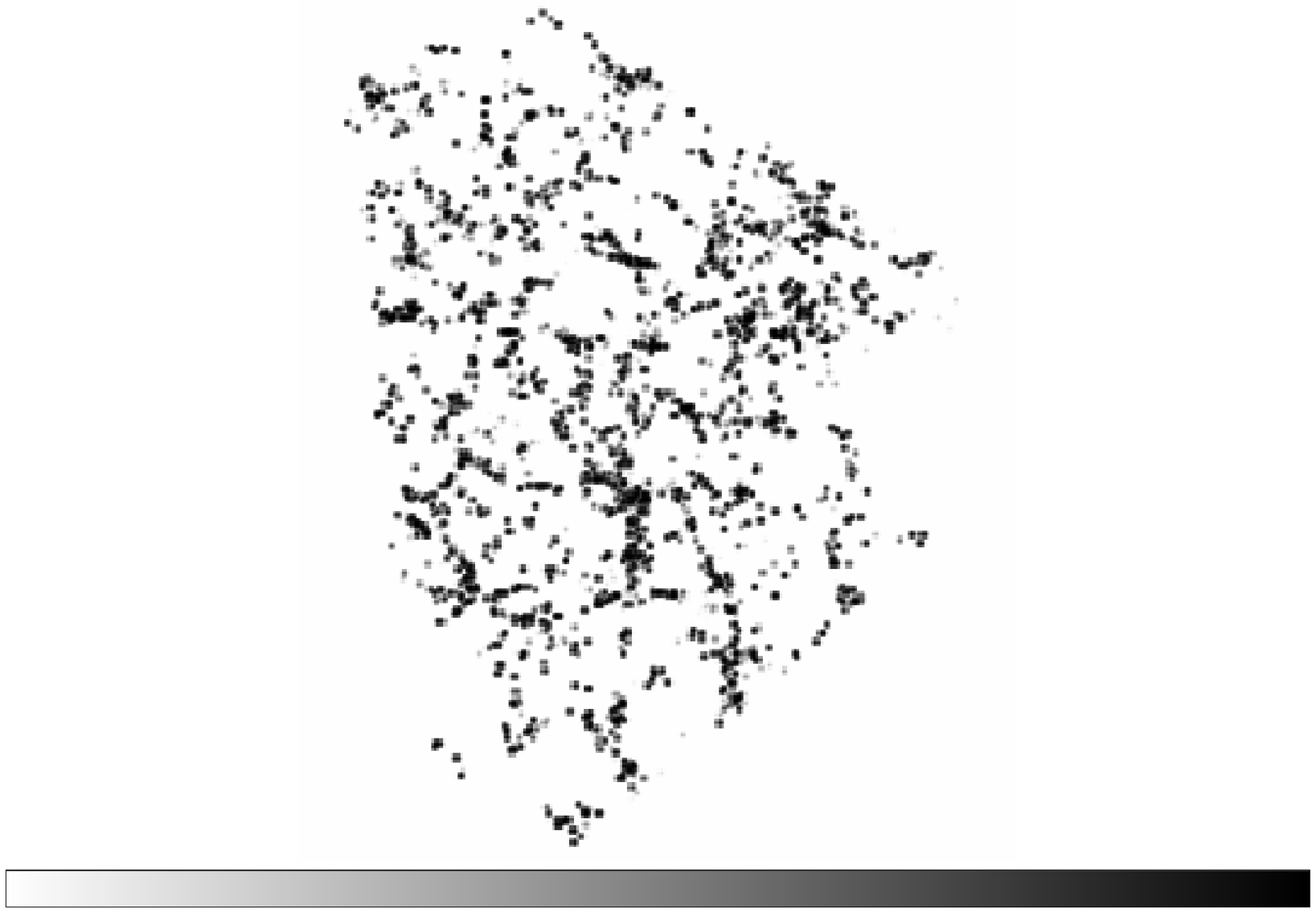}}}\\[4pt]
\fbox{\resizebox{0.25\textwidth}{!}{\includegraphics*{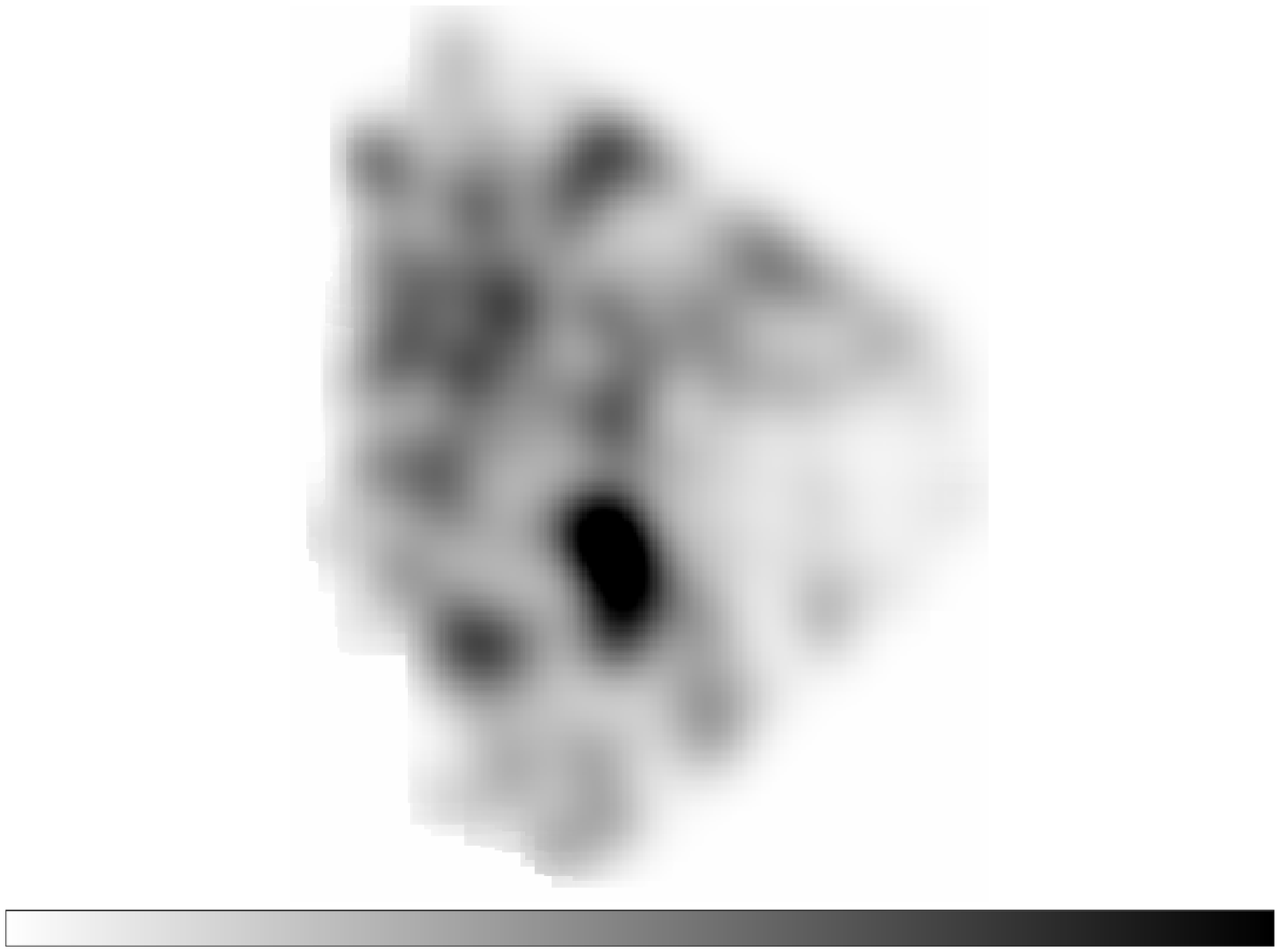}}}~
\fbox{\resizebox{0.25\textwidth}{!}{\includegraphics*{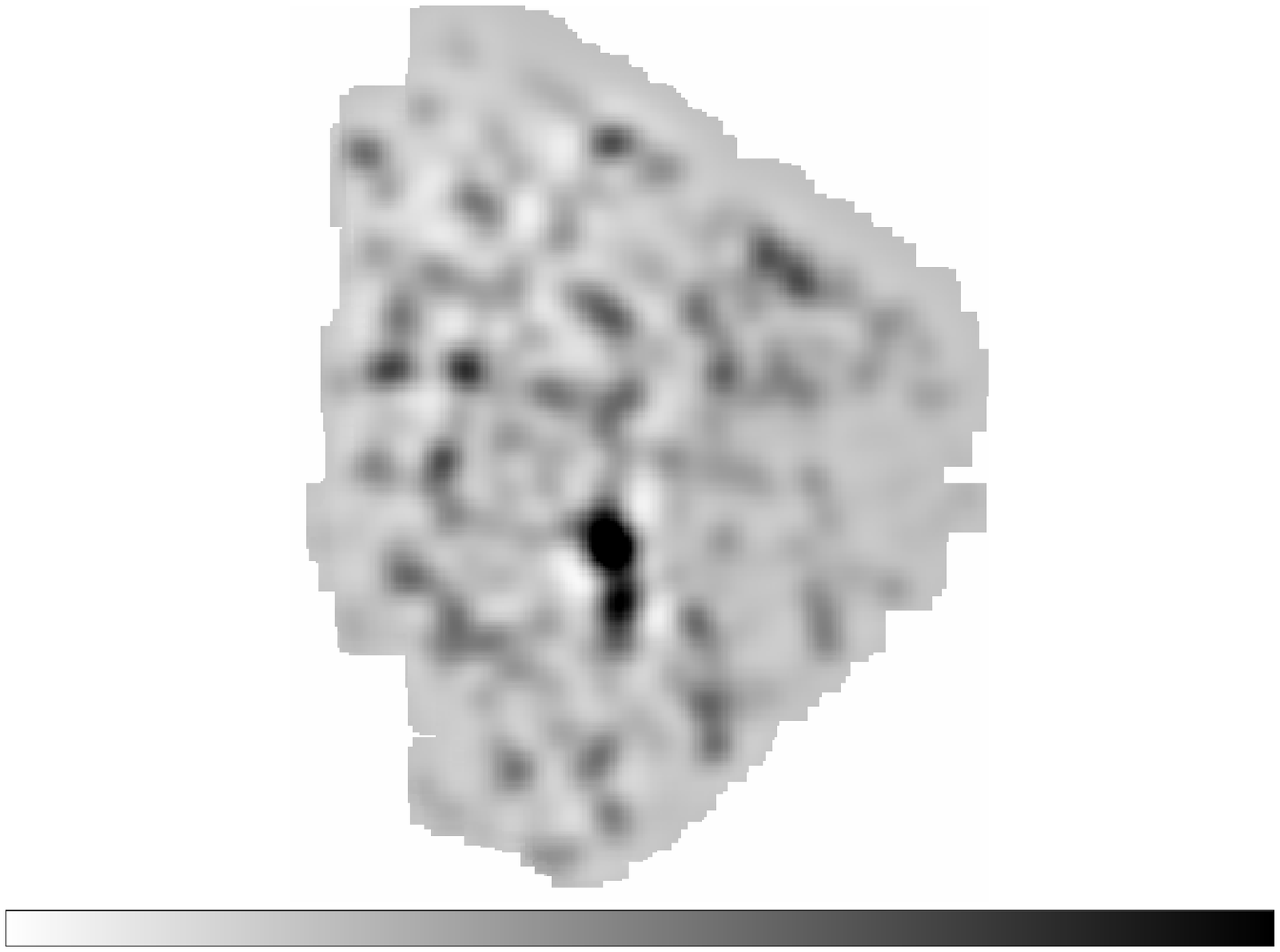}}}~
\fbox{\resizebox{0.25\textwidth}{!}{\includegraphics*{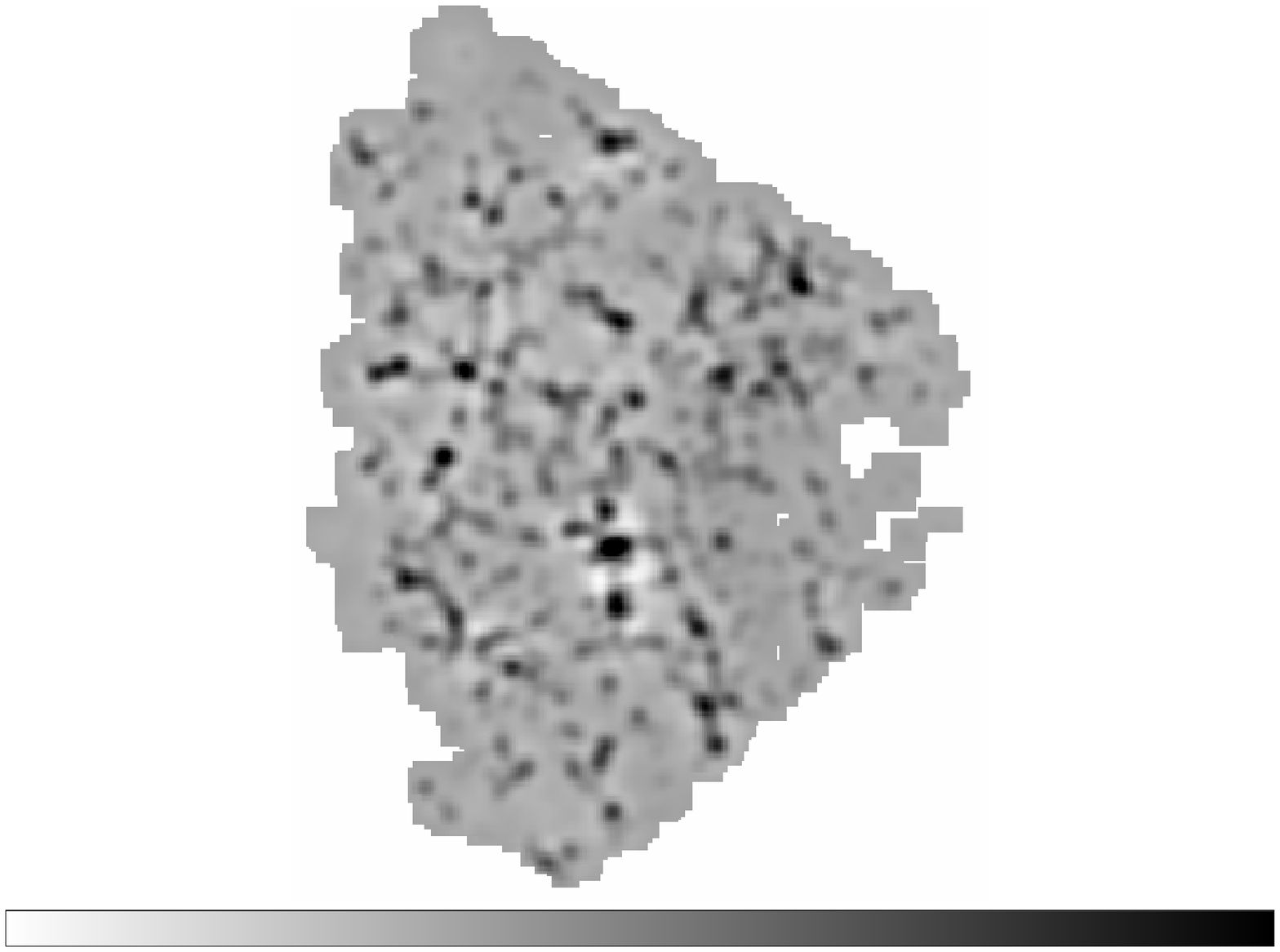}}}\\
\caption{Multi-scale decomposition of the 2df19N volume-limited sample,
for the $z=39$~Mpc/$h$ slice, and for the grid step $\sqrt2$~Mpc/$h$.
The original density is shown at top right,
the last scaling order at bottom left, the wavelet orders in between.
The wavelet orders increase from right to left and from top to bottom.
The weakest gray level shows the sample mask.
\label{fig:2df19Nbunch2}}
\end{figure*}

\begin{figure*}
\centering
\fbox{\resizebox{0.25\textwidth}{!}{\includegraphics*{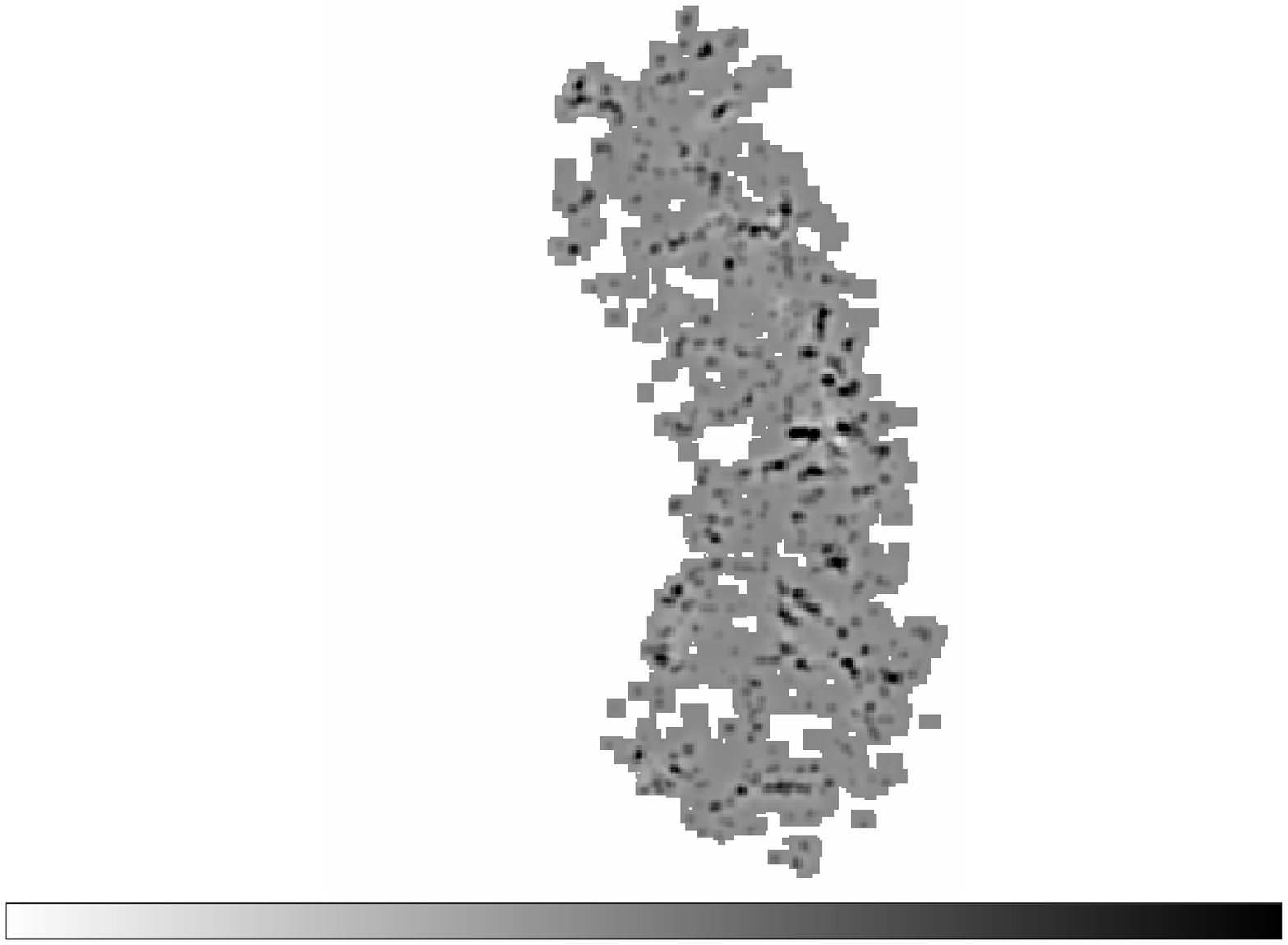}}}~
\fbox{\resizebox{0.25\textwidth}{!}{\includegraphics*{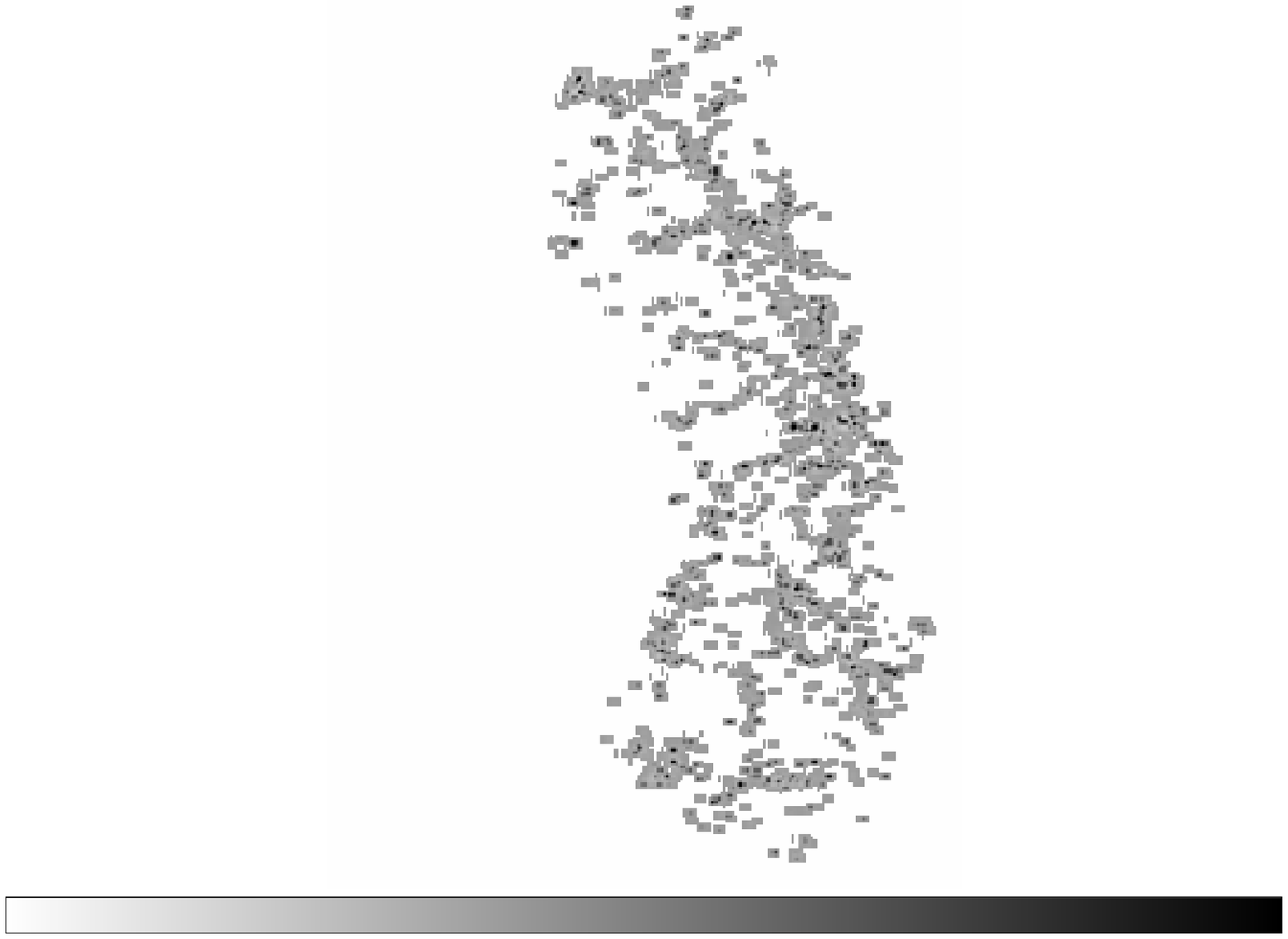}}}~
\fbox{\resizebox{0.25\textwidth}{!}{\includegraphics*{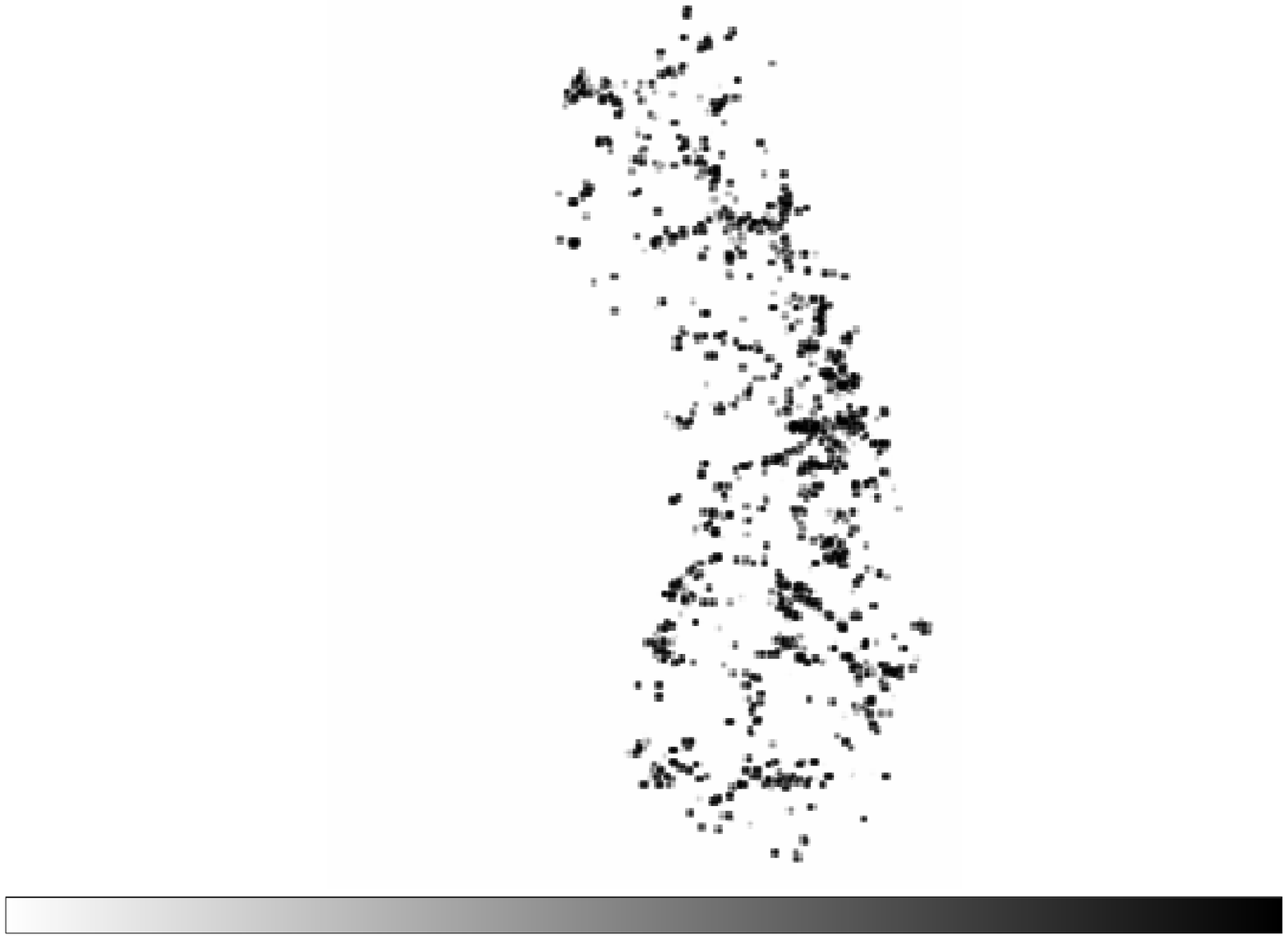}}}\\[4pt]
\fbox{\resizebox{0.25\textwidth}{!}{\includegraphics*{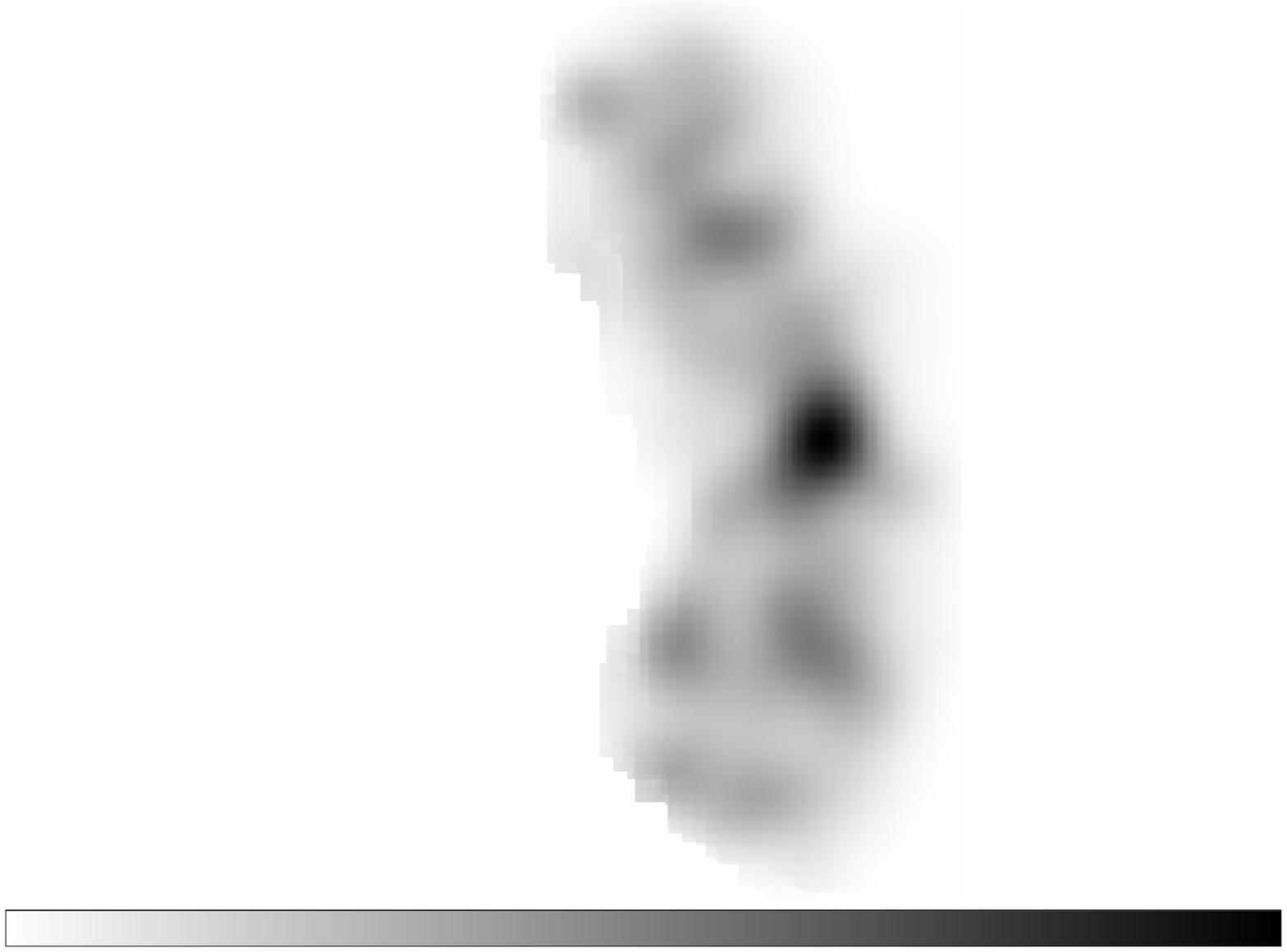}}}~
\fbox{\resizebox{0.25\textwidth}{!}{\includegraphics*{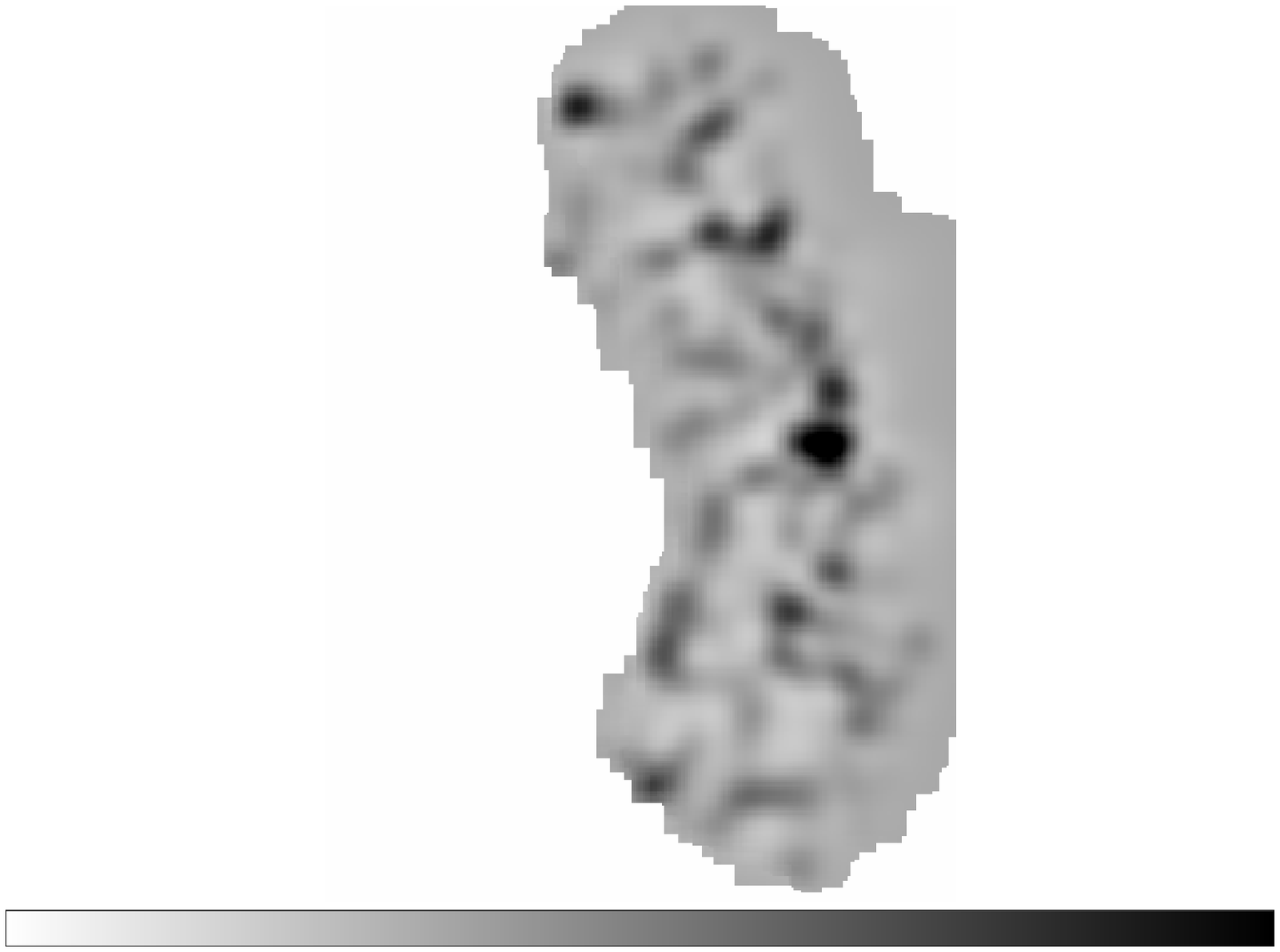}}}~
\fbox{\resizebox{0.25\textwidth}{!}{\includegraphics*{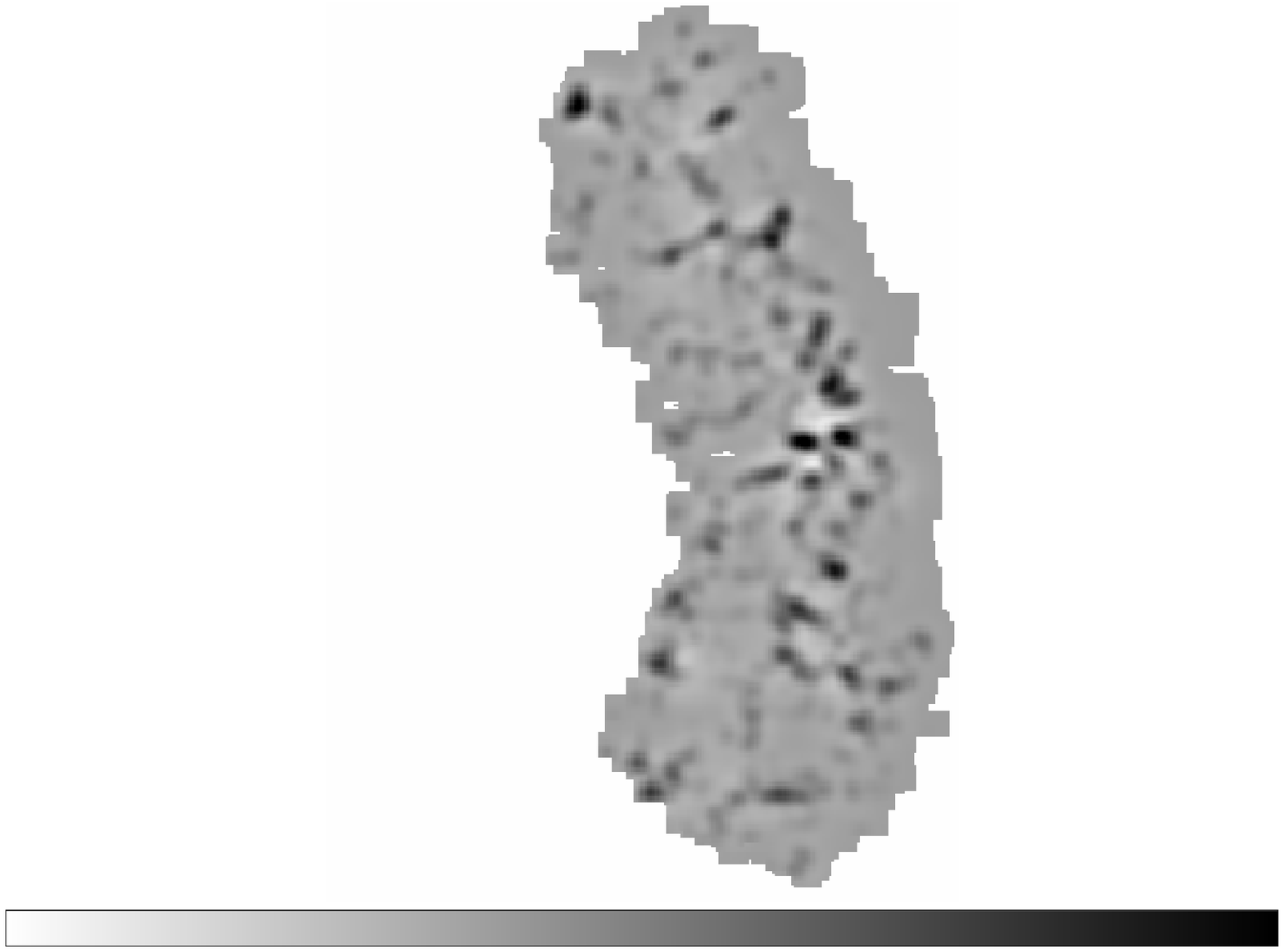}}}\\
\caption{Multi-scale decomposition of the 2df19S volume-limited sample,
for the $z=97$~Mpc/$h$ slice, and for the grid step $\sqrt2$~Mpc/$h$.
The original density is shown at top right,
the last scaling order at bottom left, the wavelet orders in between.
The wavelet orders increase from right to left and from top to bottom.
The weakest gray level shows the sample mask (in most cases; sometimes
the mask is missing and sometimes the gray level is wider than the
mask).
\label{fig:2df19Sbunch2}}
\end{figure*}

We have seen that the wavelet components of the multi-scale
decomposition of a realization of a Gaussian random field remain
practically Gaussian for simple sample boundaries. This means that
testing for Gaussianity is straightforward. However, we have to
assess the boundary effect for our application, where the boundaries
are extremely complex. We shall demonstrate it on the example of
the Northern mask. For that, we generated a realization of a
Gaussian random field for a volume encompassing the mask, as
described in section~\ref{sec:data}, and masked out the region outside
the NGP data volume. As we want to see the effects that could show
up in the data, we used the standard dark matter power spectrum for
the $\Lambda$CDM cosmology \citep{klypin97}, for the cosmological
parameters $\Omega_0=0.3, \Omega_{\Lambda}=0.7,
\Omega_{\mbox{\scriptsize bar}}=0.026, h=0.7$ (this is pretty close
to the standard 'concordance' power spectrum), generated the
realization on a grid with the step of 1~Mpc/$h$, and smoothed the
field with a Gaussian of $\sigma=2$~Mpc/$h$. The original density
distribution, the scaling distributions and the wavelets are shown
in Figs.~\ref{fig:Ngsbunch1} and \ref{fig:Ngsbunch2} for a slice at
$z=34$~Mpc/$h$, at about the middle of the sample volume.

The Minkowski functional $V_3$ for the wavelets is shown in
Fig.~\ref{fig:mf3wng}. In order not to overcrowd the figure, we do
not show the theoretical predictions. The functional has been
rescaled to show all functionals together in a single diagram and
scaling factors are shown in the labels.

As we see, the lower the wavelet order, the larger values of $V_3$
are obtained (this is also true for the other functionals);
this is expected, as higher orders represent
increasingly smoother details of the field.
The values of the functionals for the zero-order wavelet are
always higher than those for the full field, as it includes
only the high-resolution details that the iso-levels have
to follow.
Also, we have seen that the lower the order of the functional,
the smaller are the distortions from Gaussianity.

The distortions of the third Minkowski functional are the largest.
The functional for the zeroth-order wavelet (curve W0 in
Fig.~\ref{fig:mf3wng}) shows argument compression, the result of
insufficient spatial resolution of the smallest details of the
field. The functional for the first-order wavelet is close to
Gaussian, as is the functional for the total realization, but the
second-order curve 8W2 shows strong distortions, due to a small
number of independent resolution elements in the volume, and to the
small height of the slice. The characteristic volume of these
elements is $(4 \times 2^2)^3=4096$~Mpc$^3/h^3$, and their number is
about 670. The functional for the third wavelet order is already
completely dominated by noise.

So, we can estimate Minkowski functionals with a high precision, and
the border corrections work well. The most difficult part at the
moment is the scale separation in observed samples. The sample
geometries are yet slice-like, limiting the range of useful scales
by the mean thickness of the slice. Complex sample borders are also
a nuisance when applying the wavelet cascade.  In order to take
account of these difficulties, we have yet to resort to running
Monte-Carlo simulations of Gaussian realizations of a right power
spectrum, and to compare the obtained distributions of the
functionals with the functionals for the galaxy data. We hope that
for the future surveys (e.g., the full SDSS), the data volume will
be large enough to do without Monte-Carlo runs.

\section{Morphology of the 2dF19 sample}
\label{sec:datamorph}

Having developed all necessary tools, we apply them to the 2dF19
volume-limited sample, separately for the NGC and SGC regions.

We show selected slices for the two subsamples, first
(Figs.~\ref{fig:2df19Nbunch2}--\ref{fig:2df19Sbunch2}). The Northern
slice was chosen to show the richest super-cluster in the 2dfGRS
NGC, super-cluster 126 (middle and low, \citep{maret97}). The
Southern slice has the maximum area in the $z=\mbox{const}$ slices
of this sample. We choose the gray levels to show also the mask
area.

\begin{figure*}
\centering
\resizebox{.45\textwidth}{!}{\includegraphics*{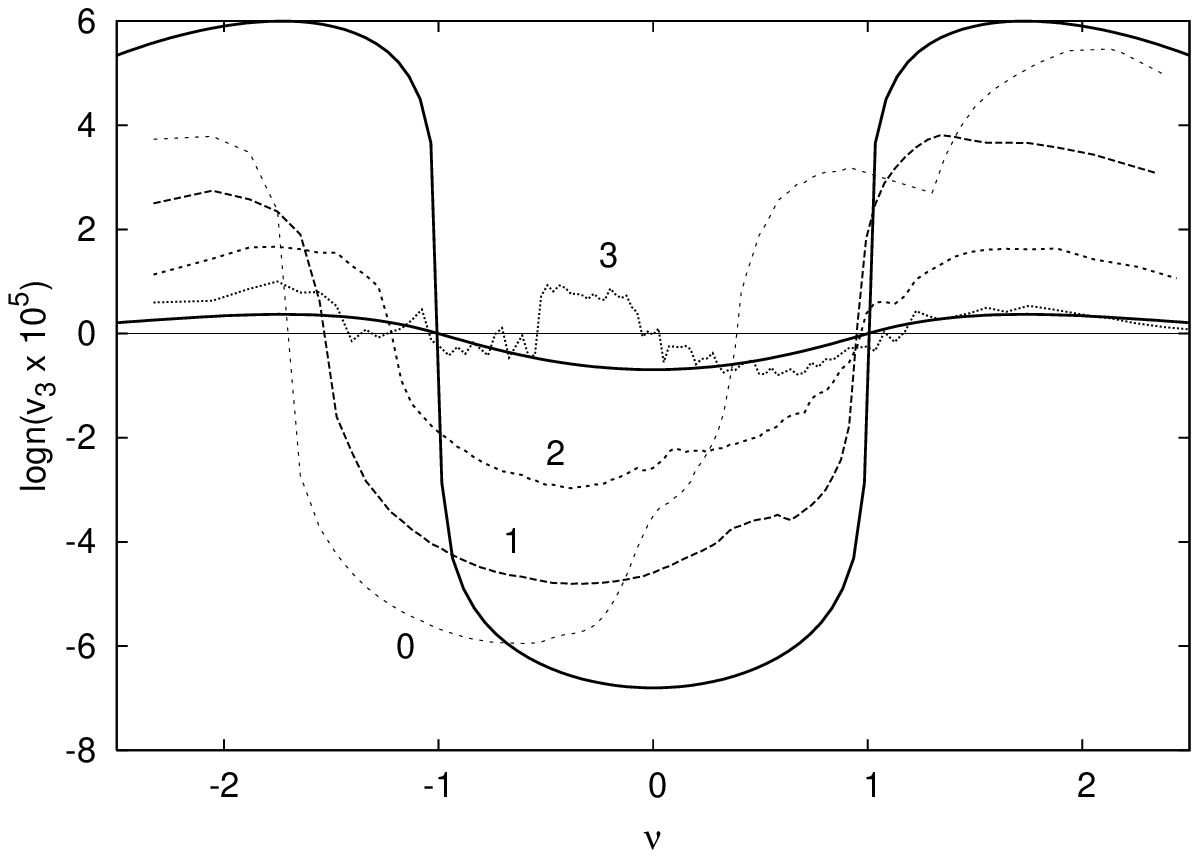}}
\resizebox{.45\textwidth}{!}{\includegraphics*{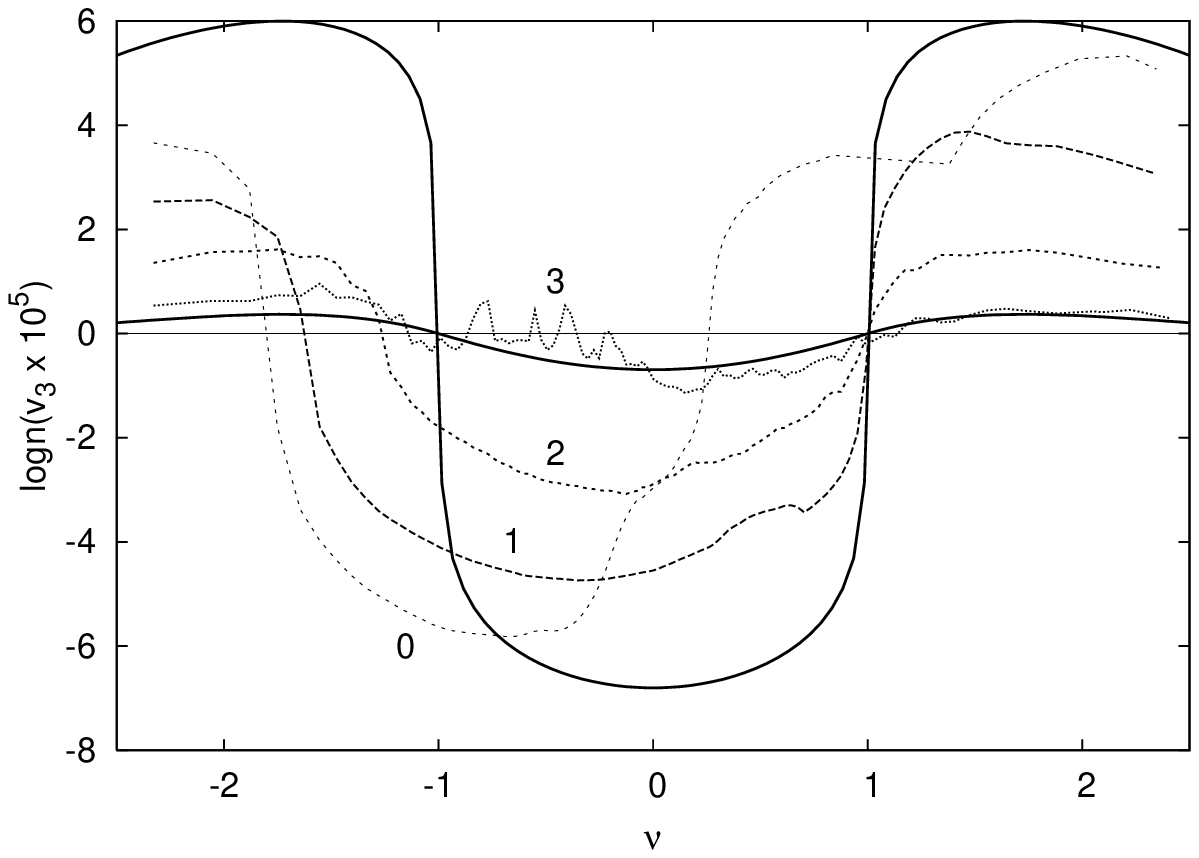}}
\caption{Summary of the densities of the fourth MF $v_3$ for the
data and all wavelet orders for the 2dFN19 sample (left) and
2dFS19 sample (right), in the logn
mapping -- the higher the wavelet order (indicated by labels), the lower the density
amplitude. Thick lines show reference Gaussian predictions. Thin
lines stand for the $\sqrt2$~Mpc/$h$ grid. \label{fig:2dfNmf3}}
\end{figure*}
\begin{figure}
\centering
\resizebox{.4\textwidth}{!}{\includegraphics*{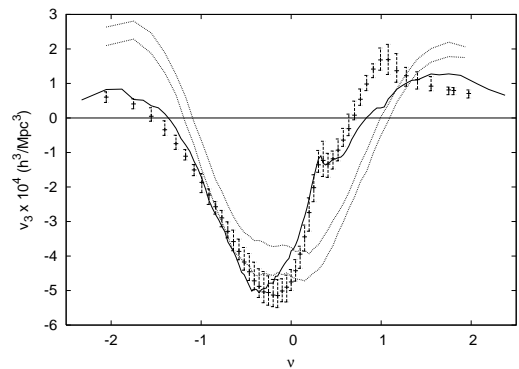}}
\caption{The density of the fourth MF $v_3$ for
the wavelet order 2 for the 2dFN19 sample (full line).
Dotted lines show the minima and maxima of 102
Gaussian realizations, and bars show the full variation in a
sample of 22 mock catalogues.
\label{fig:Nw2mf3}}
\end{figure}
\begin{figure}
\centering
\resizebox{.4\textwidth}{!}{\includegraphics*{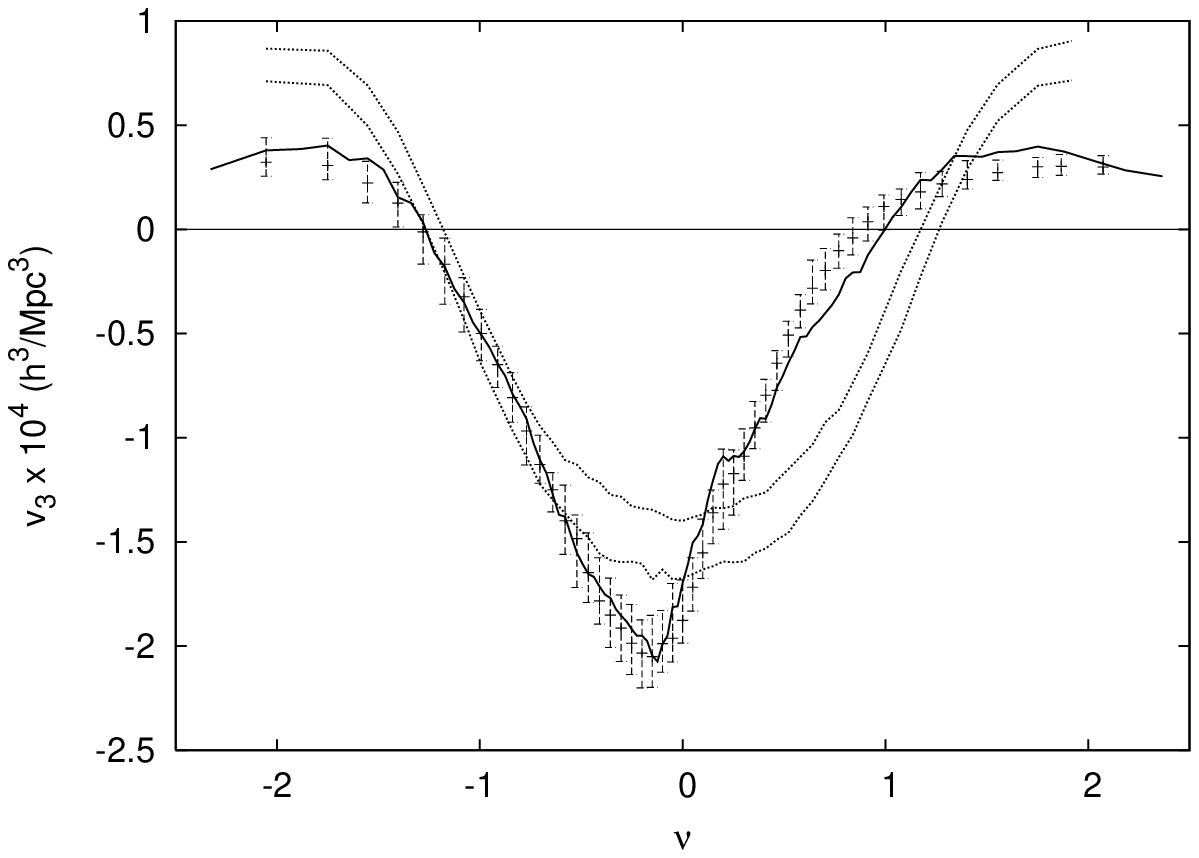}}
\caption{The density of the fourth MF $v_3$ for
the wavelet order 2 (grid $\sqrt2$) for the 2dFS19 sample (full line).
Dotted lines show the minima and maxima of 108
Gaussian realizations, and bars show the full variation in a
sample of 22 mock catalogues.
\label{fig:Sqw2mf3}}
\end{figure}
\begin{figure}
\centering
\resizebox{.4\textwidth}{!}{\includegraphics*{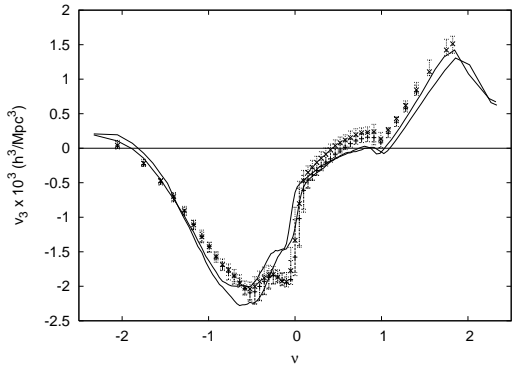}}
\caption{The densities of the fourth MF $v_3$ for
the wavelet order 1 for the 2dFS19 and 2dFN19 samples (full lines).
Bars show the full variation in two
samples of 22 mock catalogues.
\label{fig:NSw1mf3}}
\end{figure}

Figs.~\ref{fig:2dfNmf3}--\ref{fig:NSw1mf3} show the 
results -- the Minkowski functionals for
the wavelet decompositions of the 2dFN19 and 2dFS19
subsamples.

First, we show the summary figures: the density $v_3$ for the
original data and for wavelet orders from 0 to 3, and for grid unit
of $\sqrt{2}$~Mpc/$h$. Higher wavelet orders are not usable -- there
the boundary rules used for the wavelet cascade influence strongly
the results, and the number of independent resolution elements
becomes very small, letting the functionals to be dominated by
noise. The wavelet orders used span the scale range 2--22.6~Mpc/$h$.
In order to show all the densities in the same plot, we use the
mapping
\[
\mbox{logn}(v_3)=\mbox{sgn}(v_3)\log(1+|v_3|).
\]
It is almost linear for $|v_3|<1$,
logarithmic for $|v_3|>1$, and can be applied to negative
arguments, also. The density in Fig.~\ref{fig:2dfNmf3} is shown with
dotted lines. For reference, the two thick lines show the Gaussian
predictions for small and large functional density amplitudes for
the logn($v_3$) mapping. The first glance at the figures of the
fourth Minkowski functional reveals that none of the wavelet scales
shows Gaussian behavior.

In order to estimate the spread in the values of the functionals, we
ran about 100 Gaussian realizations for every sample and grid,
generated wavelets and found the Minkowski functionals. The power
spectrum for these realizations was chosen as described in
Sec.~\ref{sec:gauss} above (see \citet{klypin97}), and smoothed by a
Gaussian of $\sigma=1$ (in grid units). This is practically
equivalent to the $B_3$ extirpolation used to generate the observed
density on the grid. Now, if the observational MF-s lie outside the
limiting values of these realizations, we can say that the Gaussian
hypothesis is rejected with the $p$-value less than 1\%. We also
calculated the multiscale functionals for a set of 22 mock samples,
specially created for the 2dFGRS \cite{norberg02}. The mock catalogs
were extracted from the Virgo Consortium $\Lambda$CDM Hubble volume
simulation, and a biasing scheme described in \citet{cole98} was
used to populate the dark matter distribution with galaxies.

Fig.~\ref{fig:Nw2mf3} and ~\ref{fig:Sqw2mf3}  show  respectively the
densities of the fourth Minkowski functional for the Northern  and the Southern
Galactic caps. The MF density $v_3$ corresponding to the data is
plotted with  a continuous line while error bars correspond to the
total variation for mocks. The minimum and maximum limits for
Gaussian realizations are plotted with dotted lines.

The Gaussian realizations show very small spread, and are clearly
different from the $v_3$ Minkowski Functional of  the observational
samples. Results from mocks are much closer to the data.

Fig.~\ref{fig:Nw2mf3} shows clear non-Gaussianity for the north
galactic cap  at a high confidence level. Gaussian realizations are
not much deformed by the combination of boundary conditions
(wavelets) and border corrections (functionals) effects. In this
figure we can appreciate that with respect to this MF, mocks follow
data well for smaller density iso-levels, but deviate around
$\nu=1$; we have seen similar effects before \citep{mart05}. An
interesting detail is the knee around $\nu=0.5$, seen both in the
data and in the mocks, but not for Gaussian realizations. The latter
fact tells us that it is not caused by the specific geometry of the
data sample. The grid step was 1Mpc/$h$.

Fig.~\ref{fig:Sqw2mf3}  for the Southern data shows also clear
non-Gaussianity, but no  strong features like the northern one (it
describes larger scales, as this wavelet is based on the $\sqrt{2}$
Mpc/$h$ step grid). An interesting point is that mocks follow the
data curve almost perfectly here, much better than for the Northern
sample.

Fig.~\ref{fig:NSw1mf3} shows the $v_3$ MF density at the first
wavelet scale for both north and south slices. Again, it is clearly
non-Gaussian. However, it is remarkable how well the functionals for
both volumes coincide. This shows that the border and boundary
effects are small, and we are seeing real features in the density
distribution that we have to explain. Also, the mocks follow the
data rather well. This means that the structure (and galaxy) formation
recipes used to build the mocks already implicitly
include mechanisms responsible for these features.

It is useful to compare our results with the recent careful analysis of
the topology of the SDSS galaxy distribution by \citet{park05}.
They used Gaussian kernels to find the density distribution, and as a result,
their genus curves (see, e.g., their Figs.~6 and 8) are close to those of
Gaussian fields. They describe deviations from Gaussianity by moments of
the genus curve, taken in carefully chosen $\nu$ intervals, and normalized
by corresponding Gaussian values. Our Minkowski functionals differ so much
from Gaussian templates (see Fig.~\ref{fig:2dfNmf3}) that we cannot fit a reference
Gaussian curve. The only analogue we can find is the shift of the genus curve $\Delta\nu$,
that we estimate by fitting an expression 
$v_3(\nu)=A\left((\nu-\Delta\nu)^2-1\right)\exp\left(-(\nu-\Delta\nu)^2/2\right)$
to our results. The values of the shift corroborate the visual impression of
strong differences between our results and those of \cite{park05}. While they find
that $\Delta\nu$ lies in the interval $[-0.1,0.26]$ (for the scale range 
$R_G\in[4.5,11.0]$~Mpc/$h$, their table 2), our $\Delta\nu$ assumes values between
$-1.3$ and $-0.2$, for approximately the same scale interval 
($\lambda\in[4,22.6]$~Mpc/$h$).
As the morphology of the 2dFGRS and the SDSS should not differ much, the difference
is clearly caused by different kernels, Gaussians compared to compact wavelets. 
Dropping the conventional Gaussian
kernels makes the discriminative force of the morphological tests 
considerably stronger.
 

\section{Conclusions}
\label{sec:concl}

The main results of this paper are:
\begin{enumerate}
\item We have shown how to compute the MF, taking into account
both the biases due to the discrete
grid from the Crofton method and the border effects related to
complex observational sample volumes.

\item Our experiments have shown that the multiscale MF functionals of
a Gaussian Random Field have always a Gaussian behavior, even  in
case where the field lies within complex boundaries.
 Therefore,  we have established a solid base for calculating the Minkowski functionals
for real data sets and their multi-scale decompositions.

\item We found that both the observed galaxy density fields and mocks 
show clear non-Gaussian features of the morphological descriptors
over the whole scale range we have considered. For smaller scales,
this non-Gaussianity
of the present cosmological fields should be expected, 
but it has been an elusive quality,
not detected in most of previous papers (see, e.g., \citet{hoyle02},
\citet{park05}). But even for the largest scales that the data allows us to study
(about 20 Mpc/$h$),
the density fields are yet not Gaussian. We believe that
the Gaussianity reported in the papers cited above could be just a consequence of
oversmoothing the data. This effect was clearly described in
\citet{mart05}.

\item The mocks that are generated from initial
Gaussian density perturbations by gravitational evolution and by
applying semi-analytic galaxy formation recipes, are pretty
close to the data. However, as in a previous study \citep{mart05},
we confirm a discrepancy around $\nu=1$ between the mocks and the
data. This analysis clearly shows that there are more
faint structures in the data than in the mocks,
and clusters in the mocks have a  larger intensity
than in the real data.

\end{enumerate}

\section*{Acknowledgments}

We thank Darren Croton for providing us with the 2dF volume-limited
data and explanations and suggestions on the manuscript, and Peter
Coles for discussions. We thank the anonymous referee for constructive
comments. This work has been supported by the University of Valencia
through a visiting professorship for Enn Saar, by the Spanish MCyT
project AYA2003-08739-C02-01 (including FEDER), by the National
Science Foundation  grant DMS-01-40587 (FRG), and by the Estonian
Science Foundation grant 6104.


\section*{Appendix A: The \`a trous algorithm and galaxy catalogues}
\label{sec:atrous}

A good description of the \`a trous algorithm and of
its applications to image processing in astronomy can
be found in \citet{starck:book02}. Readers interested in the mathematical
basis of the algorithm can consult \citet{mallat:book99} and
\citet{shensa92}. We give below a short summary of the algorithm
and describe the additional intricacies that arise when the
algorithm is applied to galaxy catalogues (point data).

We start with forming the initial density distribution $d_0$ on a
grid. In order to form the discrete distribution, we have to
weight the point data (extirpolate), using the scaling kernel
for the wavelet \citep{mallat:book99}.
As we shall use the $B_3$ box spline as the
scaling kernel, the extirpolation step is:
\beq
\label{eq:extirp}
d^{(0)}(\mathbf{n}_i)=\int\rho(\mathbf{x})B_3^{(3)}(\mathbf{x-n}_i)d^3x,
\eeq
where $\mathbf{n}_i\equiv(n)_i=(x_i,y_i,z_i)$ is a grid vertex,
$\rho(\mathbf{x})$ is the original density, delta-valued at
galaxy positions, and $B_3^{(3)}(\mathbf{x})$ is the direct
product of three $B_3$ splines:
\[
B_3^{(3)}(\mathbf{x})=B_3(x)B_3(y)B_3(z),
\]
where $\mathbf(x)=(x,y,z)$.
The $B_3$ spline is given by
\[
B_3(x)=\frac1{12}\left[|x-2|^3-4|x-1|^3+6|x|^3-4|x+1|^3+|x+2|^3\right].
\]
As this function is zero outside the cube $[-2,2]^3$, every data
point contributes only to its immediate grid neighbourhood, and
extirpolation is fast.

The main computation cycle starts now by convoluting the data
$d$ with a specially chosen discrete filter $h_{(k)}$:
\beq
\label{conv}
d^{(I+1)}_{(n)}=\sum_{(k)}h_{(k)}d^{(I)}_{(n)+2^I(k)}.
\eeq
Here $I$ stands for the convolution order (octave),
the 3-dimensional filter $h_{(k)}=h_lh_mh_n$,
$(k)=(l,m,n)$
is the direct product
of three one-dimensional filters $h_i=\{1/16, 1/4,$
$3/8, 1/4, 1/16\}$,
for $i\in [-2,2]$. Two points should be noted:
\begin{itemize}
\item As the filter is the direct product of the one-dimensional
filters, the convolution can be applied consecutively for each
coordinate, and can be done in place, with extra
memory only for a data line.
\item The data index $(n)+2^I(k)$ in the convolution shows that the
data is assessed from consecutively larger regions for further
octaves, leaving intermediate grid vertices unused. This is
equivalent to inserting zeroes in the filter for these points,
and this is where the name of the method comes from (\`a trous is
``with holes'' in French). This makes the convolution very fast,
as the number of operations does not increase when the filter
width increases.
\end{itemize}

The filter $h_i$ is satisfies the dilation equation
\[
\frac12B_3\left(\frac{x}2\right)=\sum_k h_k B_3(x-k).
\]

After we have performed the convolution (\ref{conv}), we
find the wavelet coefficients $w^{(J)}$ for the octave $J$ by
simple substraction:
\beq
\label{wave}
w^{(J)}_{(n)}=d^{(J)}_{(n)}-d^{(J+1)}_{(n)}.
\eeq
The combination of steps (\ref{conv},\ref{wave}) is
equivalent to convolution of the data with the associated
wavelet $\psi^{(3)}(x)$, where
\beq
\psi(x)=2B_3(2x)-B_3(x).
\eeq
Repeating the sequence (\ref{conv},\ref{wave}) we find
wavelet coefficients for a sequence of octaves. The number
of octaves is, evidently, limited by the grid size, and in real
applications by the geometry of the sample.

We have illustrated the wavelet cascade in the main text by application to
real galaxy samples. As our wavelet amplitudes were
obtained by subtraction, we can easily reconstruct the initial density:
\beq
\label{reconst}
d^{(0)}_{(n)}=d^{(J+1)}_{(n)}+\sum_{j=0}^{j=J} w^{(j)}_{(n)}.
\eeq
Here the upper indices show the octave, the lower indices denote grid
vertices, and $d^{(J+1)}$ is the result of the last convolution.
This formula can also be interpreted as the decomposition of
the original data (density field) into contributions from
different scales -- the wavelet octaves describe contributions
from a limited (dyadic) range of scales.

Here we have to note that while the scaling kernel
\[
\Phi(x,y,z)=B_3(x)B_3(y)B_3(z)
\]
is a direct
product of three one-dimensional functions, it is surprisingly
almost isotropic. Its innermost iso-levels are slightly
concave, and outer iso-levels tend to be cubic, but this
happens at very low function values.
In order to characterise the deviation from anisotropy, let us
first define the angle-averaged scaling kernel
\[
\bar{\Phi}(r)=\frac1{4\pi}\int_S(r)\Phi(r,\theta,\phi)dS,
\]
where $S(r)$ is a spherical surface of a radius $r$.
The anisotropy can now be calculated as the
integral of the absolute value of the difference between the
kernel and its angle-averaged value:
\[
\int_2^2\int_2^2\int_2^2\left|\Phi(x,y,z)
    -\bar{\Phi}(\sqrt{x^2+y^2+z^2})\right|\,dx\,dy\,dz=0.030.
\]
As the integral of the kernel itself is unity, the deviation is
only a couple of per cent.

A similar integral over the wavelet profile gives the value
0.052. Here the natural scale, the integral of the square of the wavelet
profile, is also unity, so the deviation from isotropy
is small.

Isotropy of the wavelet is essential, if we want to be sure that our results
do not depend on the orientation of the grid. This is usually
assumed, but with a different choice of the scaling kernel
this could easily happen.

As our wavelet transform is not orthogonal, there remain correlations
between wavelet amplitudes of different octaves. The Fourier transform
of the $B_3$ scaling function is
\[
\hat{B}_3(\omega)=\left(\frac{\sin(\omega/2)}{\omega/2}\right)^4
\]
The Fourier transform of the associated wavelet (\ref{wave}) is
\[
\hat{w}(\omega)=\hat{B}_3(\omega/2)-\hat{B}_3(\omega).
\]

\begin{figure}
\centering
\resizebox{0.4\textwidth}{!}{\includegraphics*{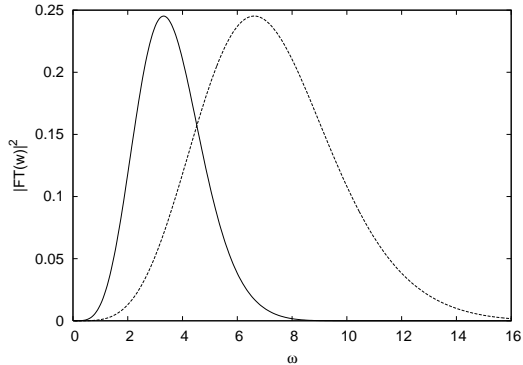}}
\caption{The square of the Fourier transform of the wavelet for
two neighbouring octaves.
\label{fig:wavefour}}
\end{figure}
We show the square of the Fourier transform of the wavelet for
two neighbouring octaves in Fig.~\ref{fig:wavefour}. As we see,
the overlap between the octaves is not large, but substantial.
This is the price we pay for keeping the wavelet transform shift
invariant. We can now compare wavelet amplitudes for different
octaves (scale ranges) at any grid vertex, but we have to
keep in mind that the separation of scales is not complete.
It may seem an unpleasant restriction that the wavelet scales have
to increase in dyadic steps. It is not, in fact, as one can choose
the starting scale (the step of the grid) at will.

Before applying the wavelet transform to the data, we have to
decide how to calculate the convolution (\ref{conv}) near the
spatial boundaries of the sample. Exact convolution can be carried
out only for periodic test data, and spatially limited data need
special consideration. For density estimation, a useful method to
deal with boundaries is to renormalise the kernel. This cannot be
done here, as renormalisation would destroy the wavelet nature
of our convolution cascade. The only assumptions that can be used
are those about the behaviour of the density outside the boundaries
of the sample. Let us consider, for example, the one-dimensional case
and the data $d(i)$ known only for the grid indices $i>0$. The possible
boundary conditions are, then:
\[
d(i;i<0)=\left\{ \begin{array}{r@{\quad:\quad}l}
    0&\mbox{zero boundary}\\
    d(0)&\mbox{constant boundary}\\
    d(-i)&\mbox{reflecting boundary}.
    \end{array}\right.
\]
The constant boundary condition is rarely used; the most popular
case seems to be the reflecting boundary. For brick-type sample
geometries, where the coordinate lines are perpendicular to the
sample boundary, this condition gives good results. However, in our
case the sample boundary has a complex geometry, and reflections from
nearby boundary surface details would soon interfere with each other.

\section*{Appendix B. Comparing border corrections}
\label{app:boundcorr}
\begin{figure}
\centering
\resizebox{0.4\textwidth}{!}{\includegraphics*{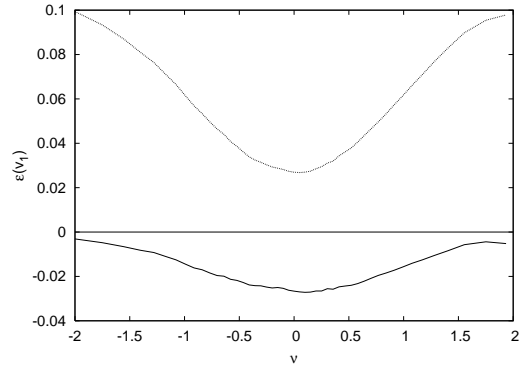}}
\caption{Relative errors of border-corrected densities of the
second MF $v_1$ for a realization of a Gaussian
random field in the 2df19N sample mask. The case of the border
correction chain is shown by solid line, the 'raw' correction case --
by dotted line.
\label{fig:cmdiffv1}}
\end{figure}

We noted above (Sec.~\ref{subsec:bound}) that alongside with
the border correction chain (\ref{corrchain}) there is
another possibility to correct for borders.
In this case we ignore the vertices in
the mask and do not build any basic elements if one of the vertices
belong to the mask. This also means that we do not have to
build the basic elements in the mask region. The latter fact makes the
algorithm faster (about twice faster for the 2dF data), as
the data region occupies usually only a fraction of the encompassing brick).

\begin{figure}
\centering
\resizebox{0.4\textwidth}{!}{\includegraphics*{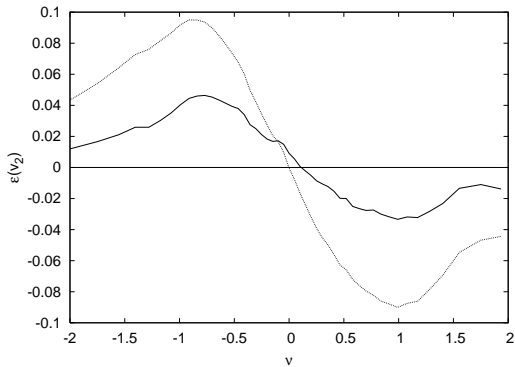}}
\caption{Relative errors of border-corrected densities of the
third MF $v_2$ for a realization of a Gaussian
random field in the 2df19N sample mask. The case of the border
correction chain is shown by solid line, the 'raw' correction case --
by dotted line.
\label{fig:cmdiffv2}}
\end{figure}

We call this method the 'raw' border correction and compare
it with the border-correction algorithm (\ref{corrchain})
on the example of the realization of a Gaussian random field
for the 2dF NGC region, smoothed by a Gaussian of $\sigma=3$~Mpc/$h$
to ensure that we resolve the density distribution. (We used the same
realization to compare the border-corrected and uncorrected case,
in Sec.~\ref{subsec:bound}.) We combine the encompassing Gaussian brick
with the 2dFN19 mask, calculate the Minkowski functionals for both
border correction methods, and compare them with the functionals
found for the periodic brick. We show below the relative errors
of the functionals, defined as
\[
\varepsilon(v_i(\nu))=\frac{v_i(\nu)-v^b_i(\nu)}{\max_\nu|v^b_i(\nu)|},
\]
where $v^b_i(\nu)$ are the densities of the functionals for the brick.
We cannot use $v^b_i(\nu)$ themselves to normalise the errors, as
their values pass through zero, so we use their maximum absolute
values.

\begin{figure}
\centering
\resizebox{0.4\textwidth}{!}{\includegraphics*{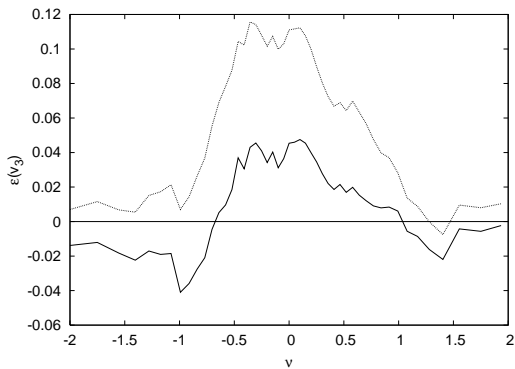}}
\caption{Relative errors of border-corrected densities of the
fourth MF $v_3$ for a realization of a Gaussian
random field in the 2df19N sample mask. The case of the border
correction chain is shown by solid line, the 'raw' correction case --
by dotted line.
\label{fig:cmdiffv3}}
\end{figure}

The relative differences are shown in
Figs.~\ref{fig:cmdiffv1}-\ref{fig:cmdiffv3}. We see that both
correcting methods give the results that do not differ from the true
densities more than 10\% (12\% for $v_3$). We see also that the
border correction chain (\ref{corrchain}) gives always better estimates of the
functionals; the maximum error is 3--4\%, and the error is about three times
smaller than that for the 'raw' border correction. Thus we use
this chain throughout the paper.

It is useful to recall, though, that the border corrections
(\ref{corrchain}) are based on the assumption of homogeneity and
isotropy of the data, which may not always be the case. The 'raw'
border corrections do not rely on any assumptions, and are therefore
useful for verifying the results obtained by the correction chain.
\label{lastpage}

\end{document}